\documentclass[fleqn]{2020SCGE}
\usepackage{color}
\usepackage{float}
\usepackage{natbib}
\usepackage{gensymb}
\usepackage{graphicx}
\usepackage{tablefootnote}
\usepackage{url}
\usepackage{hyperref}

\newcommand{\hii}{\mbox{H\,\textsc{ii}}}

\newcommand{\hi}{\mbox{H\,\textsc{i}}}
\def\degr{\hbox{$^\circ$}}

\def\fdg{\hbox{$.\!\!^\circ$}}
\def\farcm{\hbox{$.\mkern-4mu^\prime$}}

\begin{document}
\ensubject{subject}
\ArticleType{Article}
\SpecialTopic{Special Topic: Peering into the Milky Way by FAST}
\Year{2022}
\Month{December}
\Vol{65}
\No{12}
\DOI{10.1007/s11433-022-2039-8}
\ArtNo{129703}
\ReceiveDate{Jul 14, 2022}
\AcceptDate{Nov 16, 2022}

\title{Peering into the Milky Way by FAST: \\ II. Ionized gas in the
  inner Galactic disk revealed by the piggyback line observations of
  the FAST GPPS survey\footnote{\href{https://link.springer.com/content/pdf/10.1007/s11433-022-2038-7.pdf}{\color{blue}{News and views} on this paper}}}{Peering into the Milky Way by FAST:
  II. Ionized gas in the inner Galactic disk revealed by the piggyback
  line observations of the FAST GPPS survey}

\author[1\footnote{Corresponding authors (L. G. Hou, email: lghou@nao.cas.cn; J. L. Han, email: hjl@nao.cas.cn)}]{L. G. Hou}{}
\author[1,2$^{\color{blue} \dagger}$]{J. L. Han}{}
\author[1]{Tao Hong}{}
\author[1,2]{X. Y. Gao}{}
\author[1]{Chen Wang}{}


\AuthorMark{Hou L. G., Han J. L., Hong T.} 

\AuthorCitation{L. G. Hou, J. L. Han, T. Hong, X. Y. Gao and C. Wang}

\address[{\rm1}]{National Astronomical Observatories, Chinese Academy of
     Sciences, Beijing 100101, China}
\address[{\rm2}]{School of Astronomy, University of Chinese Academy of Sciences, Beijing 100049, China}     

\abstract{As one of the major components of the interstellar medium,
  the ionized gas in our Milky Way, especially the low-density diffuse
  component, has not been extensively observed in the radio band.  The
  Galactic Plane Pulsar Snapshot (GPPS) survey covers the sky area
  within the Galactic latitude of $\pm10^\circ$ around the Galactic
  plane visible by the Five-hundred-meter Aperture Spherical radio
  Telescope (FAST), and the spectral line data are simultaneously
  recorded during the pulsar survey observations. With an integration
  time of 5 minutes for each beam, the GPPS survey project provides
  the most sensitive piggyback spectra for tens of radio recombination
  lines (RRLs) in the band of 1000$-$1500~MHz for H$n\alpha$,
  He$n\alpha$, C$n\alpha$, as well as H$n\beta$ and H$n\gamma$. We
  processed the spectral data of RRLs, and obtained a sensitive
  averaged H$n\alpha$ RRL map of a sky area of 88 square degrees in
  the inner Galaxy of 33$\degr$~$\leqslant l \leqslant$~55$\degr$ and
  $|b| \leqslant$~2$\fdg$0. The final spectral data of the H$n\alpha$
  RRLs have a spatial resolution of $\sim$3$^\prime$, a spectral
  resolution of 2.2~km~s$^{-1}$, and a typical spectral rms noise of
  0.25~mJy~beam$^{-1}$ or 6.3~mK~in main-beam brightness temperature.
  The FAST GPPS H$n\alpha$ RRL observations are sensitive down to an
  emission measure of about 200~cm$^{-6}$~pc if a 3$\sigma$ detection
  limit is required.  The new H$n\alpha$ RRL map shows complex
  structural features dominated by a number of $\hii$ regions and
  large extended diffuse ionized gas regions. We detect about 94\% of
  the known $\hii$ regions and confirm 43 {\it WISE} $\hii$ regions in
  the observed sky area. Several large $\hii$ regions or star-forming
  complexes in the distant outer Galaxy are resolved in the map of
  H$n\alpha$ RRLs. Extended RRL features of the diffuse ionized gas
  are detected. In addition, the GPPS piggyback spectral-line data
  also provide sensitive detection for other kinds of RRLs, such as
  the He$n\alpha$, C$n\alpha$, H$n\beta$ and H$n\gamma$ RRLs. The RRL
  data products of the GPPS survey will be published and updated at
  {\it \color{blue}http://zmtt.bao.ac.cn/MilkyWayFAST/}.
}%
\keywords{Key Words: surveys, ISM, HII regions, radio lines}
\PACS{95.80.+p, 98.38.-j, 98.58.Hf, 95.30.Ky }

\maketitle

\begin{multicols}{2}
\section{Introduction}           
\label{sect:intro}

Ionized gas is one of the major components of the interstellar medium
(ISM). It is widely distributed in the Milky Way, from the Galactic
center to the far outer Galaxy, making up about 20\% of the total gas
mass of the Galaxy \citep[][]{lequeux05}.
Observations of the Galactic ionized gas are crucial for understanding
many relevant astrophysics, such as the star formation and $\hii$
regions, the kinematics of ionized ISM, the electron density
distribution and the spiral structure of the Galaxy, and the gaseous
metallicity and recycling of materials in the Galactic ISM
\citep[e.g.,][]{dwbw80,hipass,hh14,ska15,ngvla18,gdigs}.

In comparison to many surveys of interstellar atomic gas
\citep[e.g.,][]{lab,gass09,thor16,ebhis16,hi4pi}, molecular gas
\citep[e.g.,][]{dht11,cohrs13,chimps16,su2019,chimps20,bene21,sedigism21} and dusts
\citep[e.g.,][]{cobe98,spitzer06,wmap11,plank14}, previous surveys for the
ionized gas (see below), especially for the low-density diffuse ionized gas
\citep[e.g.,][]{gdigs}, have to be improved in aspects of the
observation sensitivity, spatial resolution, spectral resolution, and
sky coverage.

\subsection{Observation methods for the Galactic ionized gas}

Generally, ionized gas exists in the Milky Way in three forms: $\hii$
regions, diffuse ionized gas, and hot ionized gas
\citep[e.g.,][]{ferriere01,lequeux05}. The ionization sources could be
the far-UV radiation of hot stars or other mechanisms such as
collisional ionization by interstellar shocks. The $\hii$ regions
and diffuse ionized gas can be detected over a wide range of the
electromagnetic spectrum. Hot ionized gas originating from supernova
remnants and bubbles is generally detected in the high-energy band.
In the following we concentrate on $\hii$
regions and diffuse ionized gas in the Milky Way.

Some direct and indirect methods have been adopted to observe $\hii$
regions and interstellar diffuse ionized gas: (1) optical
recombination lines such as the H$\alpha$ line
\citep[e.g.,][]{haffner99,gaustad01}; (2) fine-structure lines of the
elements of oxygen~(O), nitrogen~(N), and sulfur~(S) from UV to
infrared bands \citep[e.g.,][]{cobe91,gry92,haffner99}; (3) radio
recombination lines of hydrogen~(H), helium~(He), and carbon~(C)
\citep[e.g.,][]{mh67, ch87, hrds11}; (4) dispersion measures of
pulsars in the radio band \citep[e.g.,][]{tc93, ne2001, ymw17}; (5)
radio free-free continuum emission \citep[e.g.,][]{fc72,kc97,gao19};
(6) free-free absorption of background synchrotron continuum
\citep[e.g.,][]{su2018}. Each approach has its own advantages and
shortcomings. { Since} the H$\alpha$ line is strong, { it} can be used
to probe low-density ionized gas { and} has been surveyed throughout
the entire sky \citep[][]{gaustad01,wham03}. However, observations of
the H$\alpha$ line in the optical band suffer from severe dust
extinction. Dispersion measures of pulsars are the integration of
thermal electron density from a pulsar to us, and measurements are
limited to selected sight lines to pulsars, though the models of the
free-electron distribution in the Milky Way can be constructed
\citep[e.g.,][]{tc93,ne2001,ymw17}.
Radio continuum surveys for free-free emission can avoid
interstellar dust extinction, but the decomposition of diffuse thermal
emission from the Galactic synchrotron radio emission
\citep[][]{xu2013} needs well-calibrated measurements on many bands
and careful disentangling of data \citep{planck16}.

Radio recombination lines (RRLs) are good tracers for ionized
gas. Though the strengths of RRLs are weak in general, they have some
unique advantages: (1) RRLs in radio band almost do not suffer from
extinction, hence are suitable for measuring the ionized gas in very
distant regions of the Galactic disk; (2) modern large radio
telescopes, such as the FAST and the Green Bank Telescope (GBT),
possess a great sensitivity to detect weak RRLs with a high spatial
resolution and a good spectral resolution. If such a large telescope
is equipped with a multi-beam system or a phased array feed, the
survey efficiency could be improved significantly; (3) with
developments of broadband receiver and high-speed digital backend,
multiple-transitions of RRLs in a wide observation bandwidth can be
recorded simultaneously. After a reasonable stacking, sensitive RRL
spectra can be obtained within a reasonable integration time
\citep[e.g.,][]{bal06}. Exactly relying on these advantages, RRL
observations become a feasible approach to survey the ionized gas
throughout the Milky Way \citep[e.g.,][]{ska15,ngvla18,gdigs}.

\subsection{Previous efforts in  RRL observations}

There have been many efforts on the study of the Galactic ionized gas
through observing RRLs, either by targeted observations towards $\hii$
regions or by wide-area surveys for $\hii$ regions and diffuse ionized
gas.

\subsubsection{Targeted observations towards $\hii$ regions}

$\hii$ regions are hot gas clouds with a typical electron temperature
$T_e \sim$8000~K and an electron density $n_e \sim 10^2 -
10^4$~cm$^{-3}$. They have well-defined boundaries and are generally
distributed in interstellar space with a filling factor $\eta <$~1\%
\citep[e.g.,][]{ferriere01,lequeux05}. According to the size and
evolutionary stage, $\hii$ regions are often distinguished and named
as large $\hii$ regions, compact $\hii$ regions, ultra-compact $\hii$
regions, and hyper-compact $\hii$ regions.

Soon after \citet[][]{kar59} predicted the RRLs from hydrogen and
helium in $\hii$ regions, \citet[][]{dd67}
detected H104$\alpha$ and \citet[][]{hm65} observed H109$\alpha$ from
some Galactic $\hii$ regions.
Following these predictions and observations, many efforts were
devoted to searching for the Galactic $\hii$ regions and measure their
physical properties by using RRLs
\citep[e.g.,][]{mh67,die67,rwb70,dwbw80, wwb83,ch87,lock89,
  lock96}. Their distributions were used to delineate the Galactic
spiral structures \citep[e.g.,][]{gg76,pdd04, hhs09,hh14}.
Up to now more than 7\,000 Galactic $\hii$ regions and candidates have
been catalogued \citep[][]{wise14,arm21}, but only about 1\,000 $\hii$
regions have RRL detection as of year 2009 \citep[][]{hhs09}.
In the last decade, there have been several projects aiming to discover a
large number of $\hii$ regions by using RRLs: the GBT $\hii$ Region
Discovery Survey at 4$-$8~GHz or 8$-$10~GHz
\citep[][]{hrds,hrds11,hrds18}, the Arecibo $\hii$ Region Discovery
Survey at $X$-band \citep[][]{arecibohii12}; the Southern $\hii$
Region Discovery Survey with the Australia Telescope Compact Array at
4$-$10 GHz \citep[][]{shrds17,shrds19,shrds21}; a search of radio
continuum emission from $\hii$ region candidates with the {\it Karl
  G. Jansky} Very Large Array \citep[VLA,][]{arm21}; and the survey of
RRLs of more than 500 high-mass star-forming regions at $C$-band
(4.5$-$6.9~GHz) with the Shanghai 65-m Tianma radio telescope
\citep[][]{chen20}.

At present, more than 2\,000 $\hii$ regions have RRL
detected. However, the physical parameters (e.g., electron temperature
and density) for most of them have not been well measured. A large
number of $\hii$ region candidates \citep[$\gtrsim$ 5\,000,
  see][]{wise14,arm21} remain to be confirmed with sensitive
observations.

\subsubsection{Extensive surveys for ionized gas}
\label{surveys}

Large RRL surveys have been conducted to study the Galactic {\hii}
regions and the diffuse ionized gas.
The diffuse ionized gas exists outside the well-defined $\hii$
regions, and is widely distributed in the Galaxy, typically with $T_e \sim$8000~K,
$n_e \lesssim 1$~cm$^{-3}$ and a filling factor $\eta
\sim$~20\%$-$50\% \citep[e.g.,][]{ferriere01,lequeux05},
making up about 90\% of the total mass of Galactic ionized gas
\citep[][]{rey91}.

\begin{table*}[!t]
\centering
\renewcommand\arraystretch{0.93}
\footnotesize 
\caption{RRL transitions in the frequency range of 1000$-$1500~MHz which have been recorded by the FAST GPPS survey.}
\label{rrls}
\tabcolsep 10.3pt 
\begin{tabular*}
{0.98\textwidth}{cc|cc|cc|cc|cc}
\toprule
\hline
RRL            & Frequency & RRL           & Frequency &  RRL            & Frequency &  RRL            & Frequency &  RRL            & Frequency\\
               & (MHz)     &               & (MHz)     &                 & (MHz) &                 & (MHz) &                 & (MHz) \\
\hline
H$164\alpha$   & 1477.335 & He$164\alpha$   & 1477.937 & C$164\alpha$   & 1478.072 & H$206\beta$  & 1482.886  &  H$235\gamma$   & 1491.528 \\
H$165\alpha$   & 1450.716 & He$165\alpha$   & 1451.307 & C$165\alpha$   & 1451.440 & H$207\beta$  & 1461.600  &  H$236\gamma$   & 1472.766 \\
H$166\alpha$   & 1424.734 & He$166\alpha$   & 1425.314 & C$166\alpha$   & 1425.444 & H$208\beta$  & 1440.720  &  H$237\gamma$   & 1454.317 \\
H$167\alpha$   & 1399.368 & He$167\alpha$   & 1399.938 & C$167\alpha$   & 1400.066 & H$209\beta$  & 1420.236  &  H$238\gamma$   & 1436.175 \\
H$168\alpha$   & 1374.601 & He$168\alpha$   & 1375.161 & C$168\alpha$   & 1375.286 & H$210\beta$  & 1400.138  &  H$239\gamma$   & 1418.334 \\
H$169\alpha$   & 1350.414 & He$169\alpha$   & 1350.965 & C$169\alpha$   & 1351.088 & H$211\beta$  & 1380.417  &  H$240\gamma$   & 1400.786 \\
H$170\alpha$   & 1326.792 & He$170\alpha$   & 1327.333 & C$170\alpha$   & 1327.454 & H$212\beta$  & 1361.065  &  H$241\gamma$   & 1383.528 \\            
H$171\alpha$   & 1303.718 & He$171\alpha$   & 1304.249 & C$171\alpha$   & 1304.368 & H$213\beta$  & 1342.073  &  H$242\gamma$   & 1366.551 \\
H$172\alpha$   & 1281.175 & He$172\alpha$   & 1281.697 & C$172\alpha$   & 1281.814 & H$214\beta$  & 1323.433  &  H$243\gamma$   & 1349.851 \\
H$173\alpha$   & 1259.150 & He$173\alpha$   & 1259.663 & C$173\alpha$   & 1259.778 & H$215\beta$  & 1305.136  &  H$244\gamma$   & 1333.422 \\
H$174\alpha$   & 1237.626 & He$174\alpha$   & 1238.130 & C$174\alpha$   & 1238.243 & H$216\beta$  & 1287.176  &  H$245\gamma$   & 1317.259 \\
H$175\alpha$   & 1216.590 & He$175\alpha$   & 1217.086 & C$175\alpha$   & 1217.197 & H$217\beta$  & 1269.543  &  H$246\gamma$   & 1301.356 \\
H$176\alpha$   & 1196.028 & He$176\alpha$   & 1196.516 & C$176\alpha$   & 1196.625 & H$218\beta$  & 1252.231  &  H$247\gamma$   & 1285.708 \\
H$177\alpha$   & 1175.927 & He$177\alpha$   & 1176.406 & C$177\alpha$   & 1176.514 & H$219\beta$  & 1235.232  &  H$248\gamma$   & 1270.310 \\
H$178\alpha$   & 1156.274 & He$178\alpha$   & 1156.745 & C$178\alpha$   & 1156.851 & H$220\beta$  & 1218.539  &  H$249\gamma$   & 1255.156 \\
H$179\alpha$   & 1137.056 & He$179\alpha$   & 1137.520 & C$179\alpha$   & 1137.624 & H$221\beta$  & 1202.146  &  H$250\gamma$   & 1240.243 \\
H$180\alpha$   & 1118.262 & He$180\alpha$   & 1118.718 & C$180\alpha$   & 1118.820 & H$222\beta$  & 1186.046  &  H$251\gamma$   & 1225.565 \\
H$181\alpha$   & 1099.880 & He$181\alpha$   & 1100.328 & C$181\alpha$   & 1100.429 & H$223\beta$  & 1170.232  &  H$252\gamma$   & 1211.118 \\
H$182\alpha$   & 1081.898 & He$182\alpha$   & 1082.339 & C$182\alpha$   & 1082.438 & H$224\beta$  & 1154.697  &  H$253\gamma$   & 1196.897 \\
H$183\alpha$   & 1064.307 & He$183\alpha$   & 1064.740 & C$183\alpha$   & 1064.838 & H$225\beta$  & 1139.437  &  H$254\gamma$   & 1182.897 \\
H$184\alpha$   & 1047.094 & He$184\alpha$   & 1047.521 & C$184\alpha$   & 1047.617 & H$226\beta$  & 1124.444  &  H$255\gamma$   & 1169.116 \\
H$185\alpha$   & 1030.251 & He$185\alpha$   & 1030.671 & C$185\alpha$   & 1030.765 & H$227\beta$  & 1109.713  &  H$256\gamma$   & 1155.547 \\
H$186\alpha$   & 1013.767 & He$186\alpha$   & 1014.180 & C$186\alpha$   & 1014.273 & H$228\beta$  & 1095.238  &  H$257\gamma$   & 1142.188 \\
               &          &                 &          &                &          & H$229\beta$  & 1081.014  &  H$258\gamma$   & 1129.033 \\
               &          &                 &          &                &          & H$230\beta$  & 1067.036  &  H$259\gamma$   & 1116.080 \\
               &          &                 &          &                &          & H$231\beta$  & 1053.297  &  H$260\gamma$   & 1103.325 \\
               &          &                 &          &                &          & H$232\beta$  & 1039.793  &  H$261\gamma$   & 1090.763 \\
               &          &                 &          &                &          & H$233\beta$  & 1026.519  &  H$262\gamma$   & 1078.391 \\
               &          &                 &          &                &          & H$234\beta$  & 1013.470  &  H$263\gamma$   & 1066.205 \\
               &          &                 &          &                &          & H$235\beta$  & 1000.641  &  H$264\gamma$   & 1054.202 \\
               &          &                 &          &                &          &              &           &  H$265\gamma$   & 1042.379 \\
               &          &                 &          &                &          &              &           &  H$266\gamma$   & 1030.732 \\
               &          &                 &          &                &          &              &           &  H$267\gamma$   & 1019.258 \\
               &          &                 &          &                &          &              &           &  H$268\gamma$   & 1007.953 \\ 
\bottomrule
\end{tabular*}
\end{table*}

Early surveys \citep{gc72, hp76, cahc89, cahc90, azc97, roshi00} were
conducted below 1.7~GHz and focused on understanding the properties of
diffuse ionized gas.
\citet{lock76} measured the H166$\alpha$ at some selected positions in
the Galactic longitude range from 4$^\circ$ to 44$\fdg$6, and
\citet{heiles96a,heiles96b} observed the RRLs of H$n\alpha$ as well as
He$n\alpha$, C$n\alpha$, H$n\gamma$, and H$n\delta$ around 1.4~GHz
towards more than 580 positions in the Galactic longitude ranging from
$-$3$^\circ$ to 254$\fdg$4. They investigated the distribution
and physical properties of the extended low-density warm ionized
medium.

In the last decade, large radio telescopes have conducted several
modern RRL surveys.
\citet{hipass10,hipass} analyzed the H166$-$168$\alpha$ RRLs recorded
in the $\hi$ Parkes All-Sky Survey \citep[HIPASS,][]{barnes01}. The
HIPASS data of RRLs cover a Galactic longitude range of
  $-$164$^\circ \leqslant l \leqslant 52^\circ$ and a latitude range
of $|b| \leqslant 5^\circ$, with a spatial resolution of 14$\farcm$4
and a sensitivity of 6.4~mJy~beam$^{-1}$ at a velocity resolution
of~20~km~s$^{-1}$. It has the largest sky coverage of more than 2\,000
square degrees. The data can be used for the studies of large $\hii$
regions and diffuse ionized gas.
The Survey of Ionized Gas of the Galaxy Made with the Arecibo
telescope \citep[SIGGMA,][]{liu2013,siggma} observed the
H163$-$174$\alpha$ RRLs towards the Galactic plane of
$l=33^\circ-70^\circ$, and $|b| \leqslant 1\fdg5$, with a spatial
resolution of 6$^\prime$ and a sensitivity of
0.65~mJy~beam$^{-1}$ at a velocity resolution of~5.1~km~s$^{-1}$. It
is a sensitive RRL survey for discrete objects.
The $\hi$, OH, Recombination line survey of the Milky Way
\citep[THOR,][]{thor,thor16} observed the Galactic longitude range of
$l=14^\circ-67\fdg4$ and $|b| \leqslant 1\fdg25$ with the VLA in
$L$ band, including 19 H$n\alpha$ RRLs. THOR has a sensitivity of
about 3.0~mJy~beam$^{-1}$ at a velocity resolution of
10~km~s$^{-1}$. Its high spatial resolution of about 0$\farcm$67 is
most suitable for measuring the compact sources.
The GBT Diffuse Ionized Gas Survey \citep[GDIGS,][]{gdigs} is a
project to observe RRLs in the Galactic longitude range of $-5^\circ <
l < 32\fdg3$, and $|b| < 0\fdg5$ at $C$ band~(4$-$8~GHz), with some
extensions along the Galactic latitude directions for some selected
fields, such as the star-forming complex W\,43. The sensitivity of the
GDIGS is about 10.3~mK at a velocity resolution of 0.5~km~s$^{-1}$ for
the H$n\alpha$ transitions, and the spatial resolution is about
2$\farcm$65.  The GDIGS is a sensitive survey for both $\hii$ regions
and diffuse ionized gas at $C$ band.
The GLOSTAR survey \citep[][]{glostar} aims to cover the Galactic
plane in the range of $-2^\circ < l < 60^\circ$, $|b| < 1^\circ$ and
$76^\circ < l < 83\fdg3$, $-1^\circ < b < 2\fdg3$ at $C$ band
with the VLA (GLOSTAR-VLA) and the Effelsberg 100-m telescope
(GLOSTAR-Eff), including 7 and 24 RRLs in the VLA and Effelsberg
observations, respectively. The first results of the GLOSTAR survey
for the Galactic longitude range of $28^\circ < l < 36^\circ$ have been
presented by \citet{glostar}.

Obviously, the RRLs have attracted many attentions in the last few years, since
the properties of diffuse ionized gas in a large portion of the
Galactic disk have not been well observed.

\begin{figure*}
  \centering 
  \includegraphics[width=0.44\textwidth]{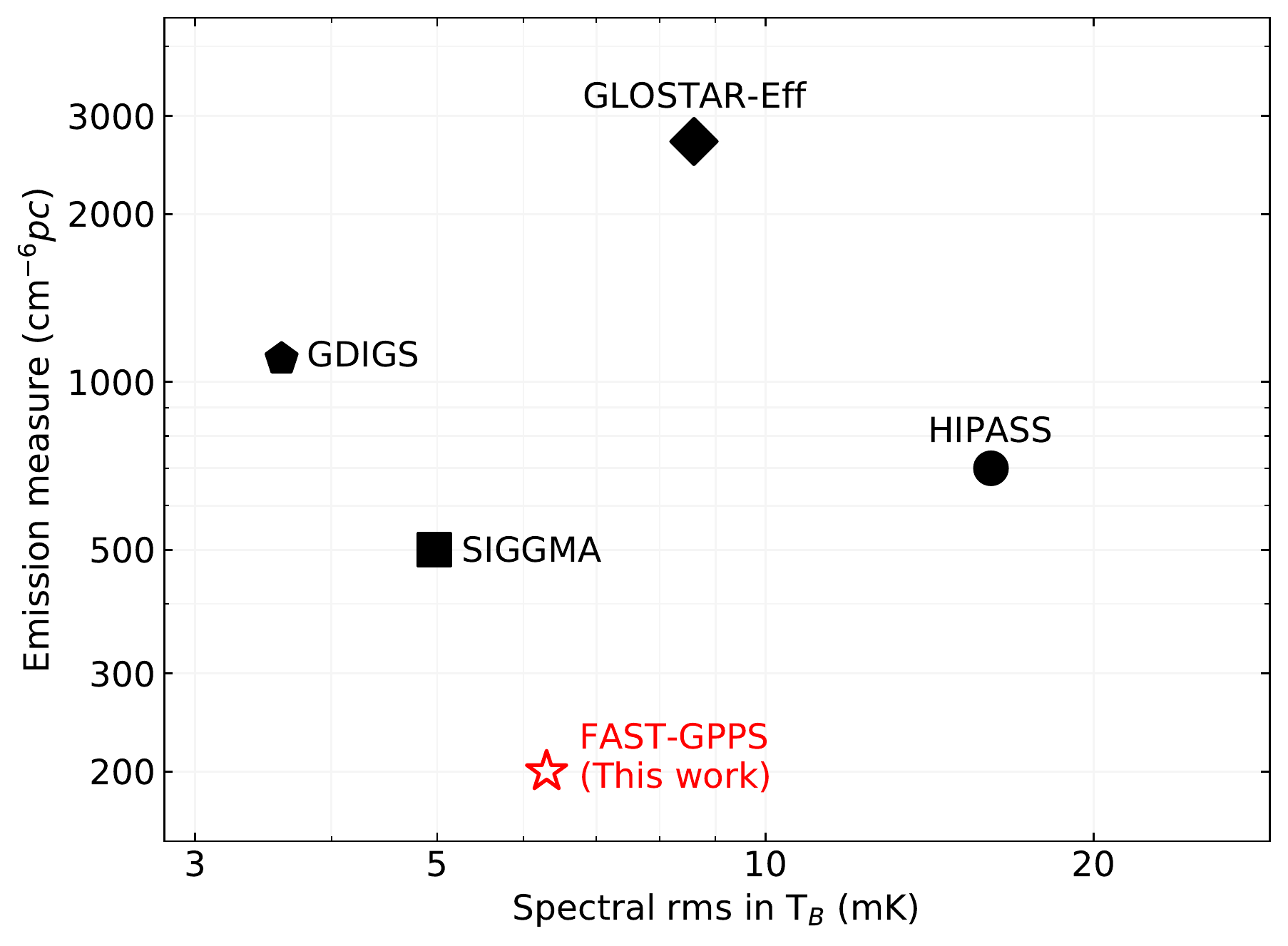}
  \includegraphics[width=0.44\textwidth]{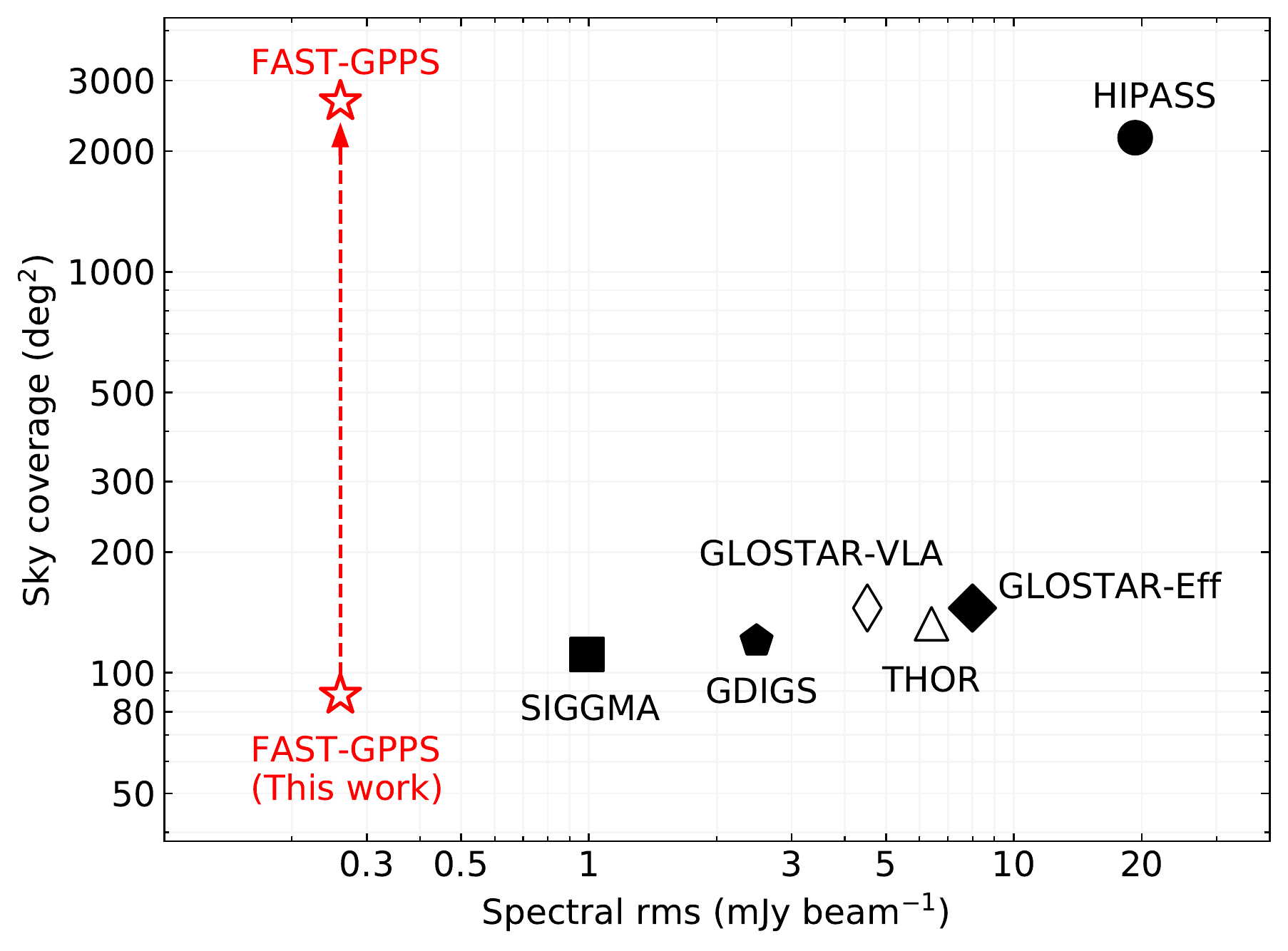} 
  \caption{{\it Left}: The sensitivity of GPPS H$n\alpha$ RRL
    observations is compared with other RRL surveys in terms of
    brightness temperature $T_B$ and the emission measure EM. The best
    surveys should have parameters in the bottom-left. The $T_B$ and
    EM values for the GLOSTAR-VLA and THOR projects are { too large
      to} show in the plot since observations are carried out by the
    synthesis radio telescope with a very high spatial
    resolution. {\it Right}: The sky coverage and the typical spectral
    rms of the FAST GPPS survey are compared to those of other RRL
    surveys. The best surveys should have parameters in the
    top-left. The spectral rms values are scaled to the same spectral
    resolution of 2.2~km~s$^{-1}$.}
   \label{comsen}
\end{figure*}

\subsection{Peering into the interstellar medium by FAST}

A systematic survey of RRLs with a high sensitivity, a good spatial
resolution and extensive sky coverage will significantly improve our
understanding on the diffuse ionized interstellar
gas~\citep[e.g.,][]{gdigs}, which is currently a major omission among
the multi-phases of the ISM.

The Five-hundred-meter Aperture Spherical radio Telescope (FAST) is
the largest single-dish telescope in the world, with an illuminated
aperture of 300-m in diameter \citep{Nan2011}. With the $L$-band
19-beam receivers covering the observational band of 1000$–$1500~MHz
\citep{fast19,fast20}, it is a powerful equipment to observe
pulsars~\citep{gpps}, $\hi$ line~\citep{Hong2022} and RRLs.
The Galactic Plane Pulsar Snapshot (GPPS) survey aims to make the most
sensitive systematic search for pulsars within the Galactic latitude
of $\pm$10$^\circ$ of the Galactic plane visible by the FAST
\citep[][]{gpps} within an angle of originally $26\fdg4$ from the
zenith but now extended to $28\fdg5$. The GPPS survey observations are
conducted by using the snapshot mode. A single snapshot contains four
nearby pointings of the 19 beams, 5-minute integration time for each
pointing, together with three quick position switches done in some
seconds. It consumes 21 minutes in total for a cover. The four
pointings of the 19 beams fully cover a sky area of 0.1575 square
degrees \citep[see Figure~4 in][]{gpps}. The GPPS survey speed can be
set to about 3 to 3.5 hours per square degree if we add the slewing
time of 10 minutes for each cover. This high efficiency usage of the
telescope time is crucial for a large-area survey.

With the digital spectroscopy backend connected to the $L$-band 19-beam
receiver, the piggyback spectral line data in the band of
1000$-$1500~MHz are recorded simultaneously during the GPPS survey. The
high sensitivity due to the huge collecting area of FAST and the 5-minute integration
time for each beam, together with an excellent spectral
resolution (channel spacing $\sim$0.477~kHz), make the 
spectral line data from the GPPS survey valuable to
reveal the characteristics of interstellar ionized gas.

This paper series is dedicated to investigations of the interstellar
medium by FAST. The first paper is presented by \citet{Hong2022} for
the piggyback $\hi$ data recorded in the GPPS survey. This is the
second paper. In the third paper by \citet{Xu2022}, the magnetic
fields in the Galactic halo and farther spiral arms are revealed by
new measurements of Faraday effect of a large sample of pulsars
observed mostly by the GPPS survey. The FAST scan observations for
radio continuum emission have been made to a sky region for the
identification of two large supernova remnants as presented by
\citet{Gao2022} in the fourth paper.

By using the piggyback spectral line data obtained during the GPPS
survey \citep[][]{gpps}, we aim to obtain the data cubes of H$n\alpha$
RRLs, and we will also work on He$n\alpha$, C$n\alpha$, H$n\beta$ and
H$n\gamma$ RRLs in the future (see Table~\ref{rrls}). The sensitivity
of the GPPS H$n\alpha$ RRL observations is compared with previous
surveys as shown in Figure~\ref{comsen}.  In this paper, we publish
the { H$n\alpha$} RRL map in the area covering $33^\circ \leqslant l
\leqslant 55^\circ$ and $|b| \leqslant 2\fdg0$. In Sect.~2, we
describe the observation status and the data processing for the RRLs
of the FAST GPPS survey. Results on RRLs are presented and discussed
in Sect.~3. The conclusions and discussions are given in Sect.~4.

\begin{figure*}[!t]
  \centering
   \includegraphics[width=0.96\textwidth]{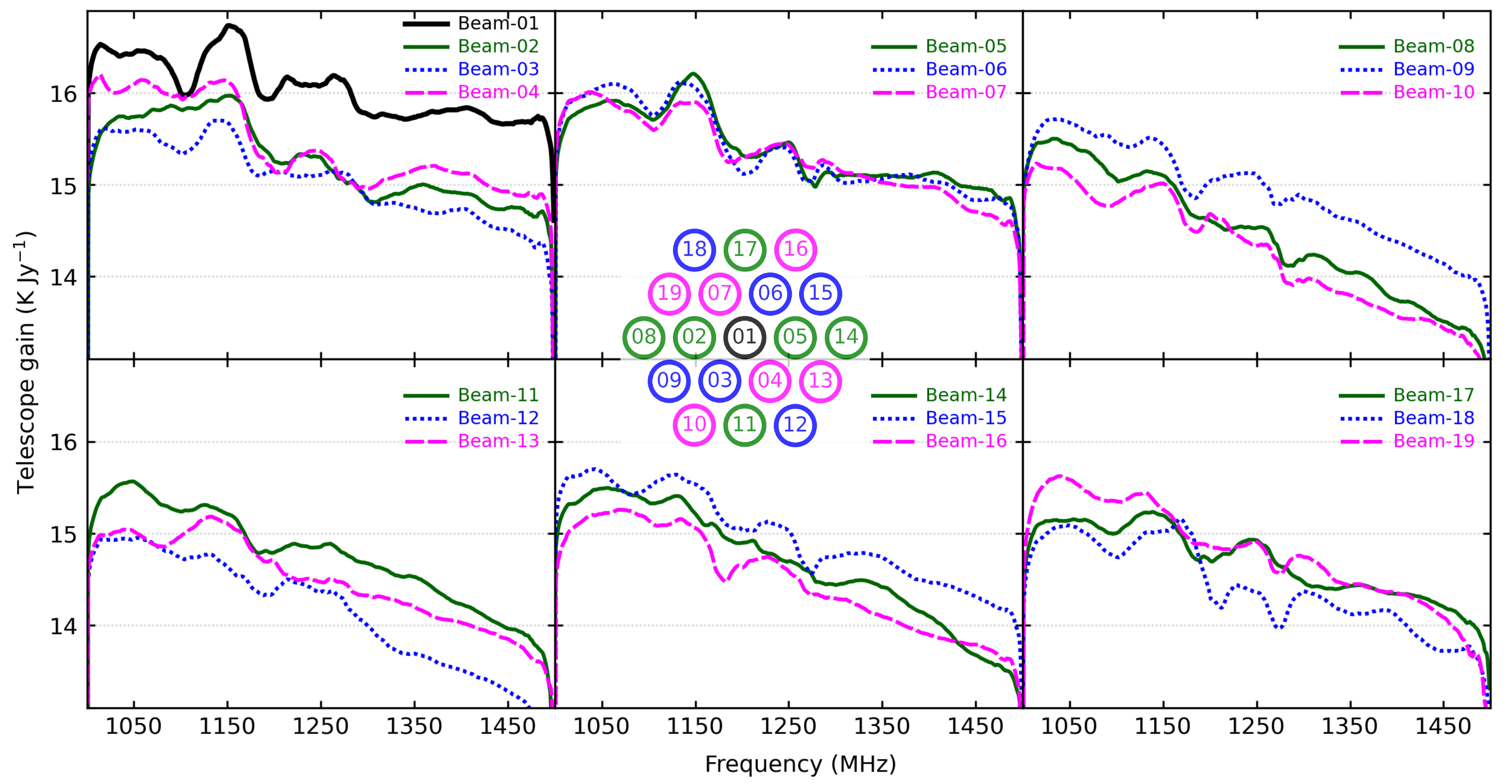}   
   \caption{The averaged telescope gain as a function of the
     observation frequency for the FAST $L$-band 19-beam receiver. The
     sky positions for 19 beams are also illuminated in the figure
     center for convenience.}
   \label{gain}
\end{figure*}

\section{Processing the RRL data of the GPPS survey}
\label{sect:Obs}

As one of five FAST key science projects approved by the FAST science
committee at the end of 2019, the GPPS survey\footnote{\it
  \color{blue}http://zmtt.bao.ac.cn/GPPS/} aims to hunter for pulsars
in all accessible sky by FAST with a Galactic latitude in the range of
$|b| < 10^\circ$, and the highest priority is given to the inner
Galaxy with $|b| < 5^\circ$ \citep[see][for details]{gpps}. The survey
formally started in February 2020. Some test observations were made in
2019.  The main observational parameters of the survey are listed in
Table~\ref{tab1}.

Observations for the GPPS survey are conducted by using the snapshot
mode, in which four adjunct pointings are finely adjusted by a step of
$3'$, the same as the size of the $L$-band receiver beams, to fully
cover a sky patch of 0.1575 square degrees, and nearby covers can also
be joined together. That is to say, the sky region so observed is
fully covered though not Nyquist sampled with the telescope beam
size. In each of the four pointings, all 19 beams track the given
positions in the sky for 5~minutes, which ensures the great
sensitivity of observations.

\begin{table}[H]
\caption{Survey parameters of the FAST GPPS survey and for the RRLs.}
\renewcommand\arraystretch{0.93}
\small
\label{tab1}
\tabcolsep 3pt 
\begin{tabular*}{0.47\textwidth}{cc}
\toprule
FAST GPPS survey           & Parameter value \\
\hline
Galactic longitude range  &    $\sim30^\circ \lesssim l \lesssim \sim98^\circ$, $\sim 148^\circ \lesssim l \lesssim \sim 216^\circ$ \\
Galactic latitude range  &    $|b| < 10^\circ$ \\
Observed frequency range    &  1000$-$1500~MHz \\
Effective frequency range   &  1050$-$1450~MHz \\
Beam size                   & $\sim3^\prime$    \\
Integration time  & 5~minutes            \\
Channel number    & 1024~K             \\
Polarization products   &  XX, YY, X$^*$Y, XY$^*$    \\
Sampling time           &  1~s               \\
\hline
This first RRL data release  &  Parameter value \\
\hline
Sky coverage  &   $33^\circ \leqslant l \leqslant 55^\circ$, $|b| \leqslant 2\fdg0$\\
 RRLs analyzed             & H183$\alpha-$H165$\alpha$      \\
LSR velocity range         & $-200$~km~s$^{-1}$ to $200$~km~s$^{-1}$ \\
 Channel spacing           & 2.2~km~s$^{-1}$   \\
Typical spectral rms noise  & 0.25~mJy~beam$^{-1}$ (6.3 mK $T_B$)        \\
\hline
\bottomrule
\end{tabular*}
\end{table}

Each observation session consists of observations for many snapshot
covers, often lasting for 2 to 4 hours. A periodical reference signal
with an amplitude of about 1~K is injected in advance of some snapshot
covers, typically before the first cover, in the middle and after the
last cover, which ensures the spectral line data can be reliably
calibrated. The gains of the FAST $L$-band 19-beam receiver (see
Figure~\ref{gain}) are stable over a few hours, typically with a
fluctuation of only a few percent \citep[][]{fast20}.

During the FAST GPPS survey, four polarization products (XX, YY,
X$^*$Y, XY$^*$) of spectral lines are simultaneously recorded for all
of the 19 beams. The accumulating time of spectral data for recording
is 1~second. The raw spectral data have 1\,024~K channels covering the
band of 1000$–$1500~MHz, which corresponds to a channel width of about
0.477~kHz or a velocity resolution of about 0.10~km~s$^{-1}$ for the
$\hi$ line at 1420~MHz. Though the system gains at the edges of the
band outside 1050~MHz and 1450~MHz are degraded~(see
Figure~\ref{gain}), the spectral data there can still be calibrated
and useful.

The GPPS piggyback RRL data will be fully processed and gradually
released, relying on the GPPS survey progress that in practice {
  depends} on the tight FAST observation schedule.

\begin{figure*}[!t]
  \centering
   \includegraphics[width=0.93\textwidth]{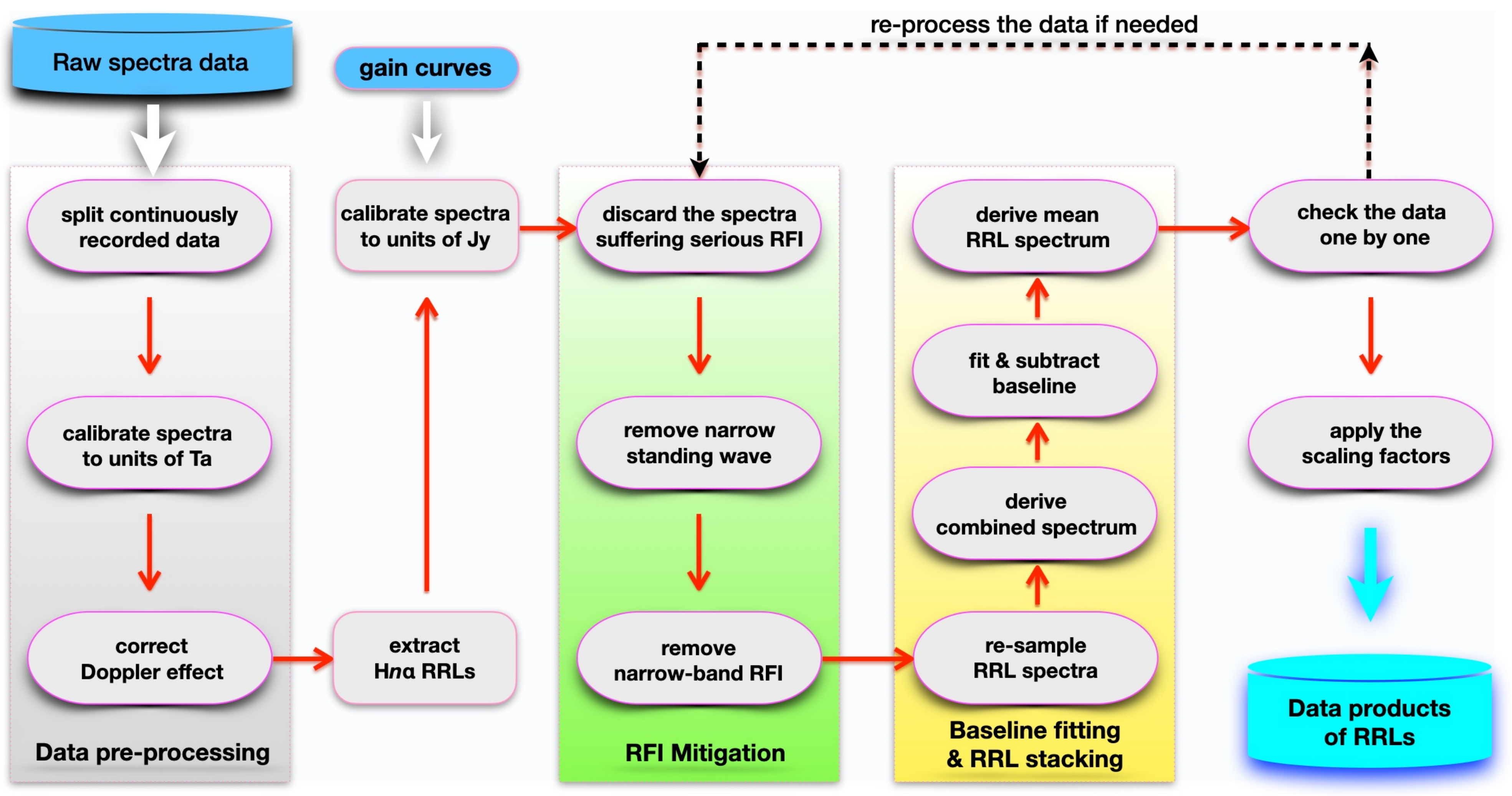}   
   \caption{The flowchart for data processing for the piggyback RRL
     data of the FAST GPPS survey.}
   \label{flowchart}
\end{figure*}

In the following, we describe the data processing for the H$n\alpha$
RRLs. The flowchart is shown in Figure~\ref{flowchart}.

\subsection{Pre-processing of the raw data}

The pre-processing of the raw spectral data acquired by the snapshot
observation generally includes three steps \citep[see][for
  details]{Hong2022}:
(1) splitting the continuously recorded data into { FITS} files for
every pointing of each beam, which contain the full polarization
information XX, YY, X$^*$Y, and XY$^*$, resulting in 76 { FITS}
files for one snapshot covering a sky area of 0.1575 square degrees;
(2) calibrating the XX and YY products to the units of antenna
temperature $T_a$ by using the scales obtained from the periodically
injected reference signal. Then, the power values of XX and YY are added to
get the spectrum of total intensity;
(3) correcting the velocity to the local standard of rest (LSR)
frame.
In this stage, the pre-processed spectral data are downgraded from
1024~K channels to 64~K channels to enhance the signal-to-noise ratio of
RRLs, which corresponds to a channel width of 7.813~kHz or a velocity
resolution of about 2.2~km~s$^{-1}$ at 1064.037~MHz (H183$\alpha$).
This velocity resolution is good enough to depict the profiles of
strong RRLs, since the typical RRL widths (full-width at half maximum,
FWHM) is about 25~km~s$^{-1}$ \citep[e.g.,][]{gdigs}.

\begin{figure*}[!t]
  \centering
   \includegraphics[width=0.92\textwidth]{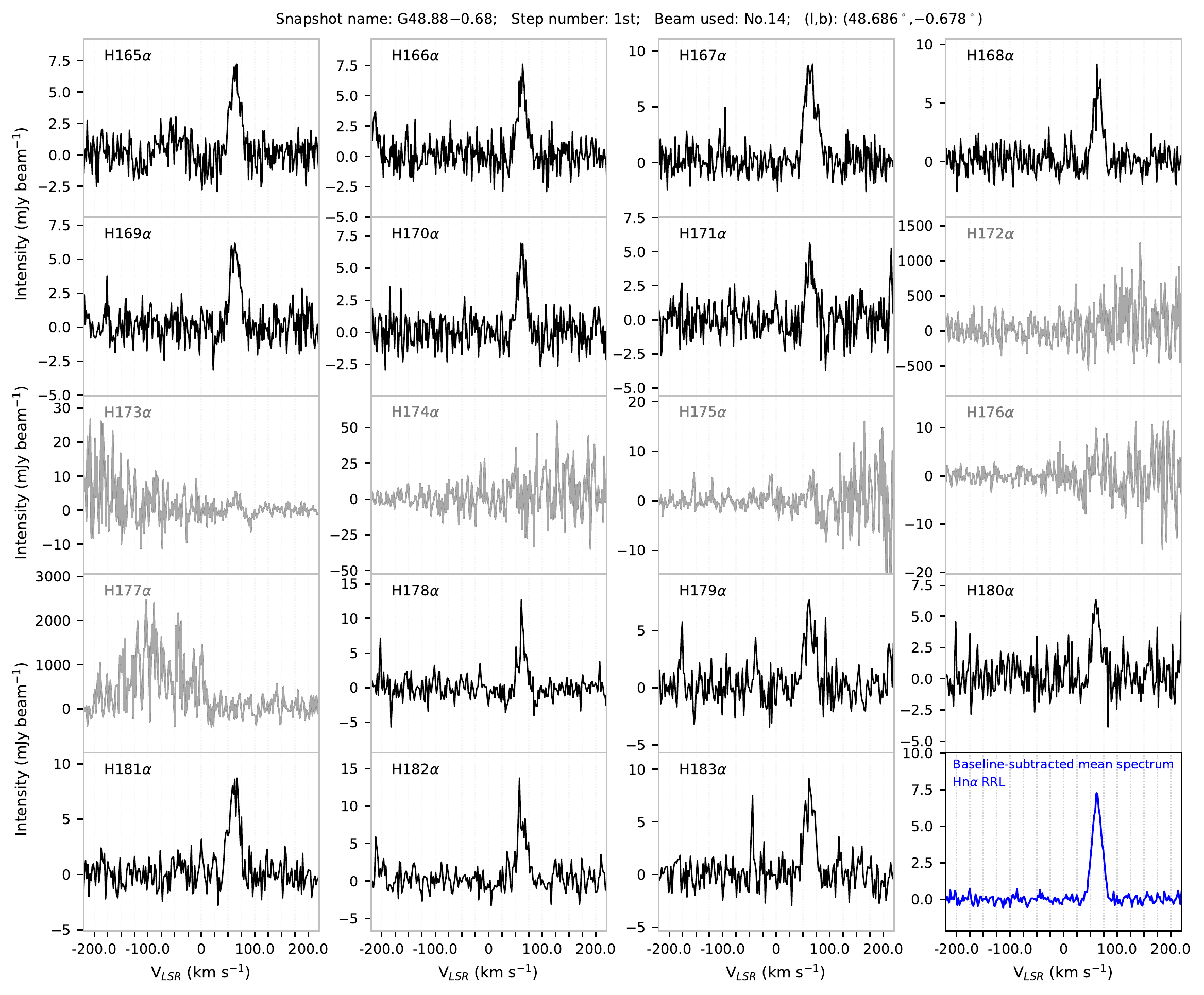}   
   \caption{An example of the H$n\alpha$ RRL
     spectra recorded by the FAST GPPS survey, specifically for the
     line data from the beam No.14 of the $L$-band 19-beam receiver of
     FAST with the 5-minute integration in the first pointing of the
     snapshot observation towards G48.88$-$0.68. Some RRL spectra
     (e.g. H172$\alpha-177\alpha$) suffering from severe RFI are not used
     in the data stacking, as indicated by grey lines. The ArPLS
     algorithm \citep{arpls,zcl+21} is adopted to fit the baseline of
     the combined spectrum. After a subtraction of the fitted
     baseline, the averaged H$n\alpha$ RRL spectrum is obtained in the {\it
       bottom right} panel.}
   \label{lines}
\end{figure*}

Subsequently, we extract 19 H$n\alpha$ RRLs from a pre-processed
spectrum, including H$183\alpha$ at a rest frequency of 1064.307~MHz
to the H$165\alpha$ at 1450.716~MHz, and convert the scale from the
units of $T_a$ to the flux density unit of Jy according to the
averaged telescope gain curves (Figure~\ref{gain}).
Based on the long-term monitoring of the standard flux density calibrators of
3C\,138 and 3C\,286, variations of the telescope gain curves are found
to be generally less than 10\% \citep[e.g.,][]{Gao2022}. In this first
data release, we adopt an averaged gain curve for each of the 19 beams
as shown in Figure~\ref{gain}, and
extract a median correction factor in the corresponding V$_{\rm LSR}$
range of $-300$~km~s$^{-1}$ to $300$~km~s$^{-1}$ for a RRL of a given
beam, then apply it to the RRL spectrum.
The calibrated RRL spectra (see examples in Figure~\ref{lines}) are used
in the following analysis.

\subsection{Mitigation of the radio frequency interference}

In some frequency channels (e.g., 1175, 1207, 1270, and 1480~MHz),
there always exist strong and broad-band ($>$~1~MHz) radio frequency
interference (RFI) signals caused by satellites and/or civil aviation
\citep[e.g.,][]{fast20,rfifast}, and some narrow-band RFI signals
occasionally emerge in some channels. 
In addition, the observed spectra are always contaminated by the
standing waves of about 1~MHz, and occasionally by the RFI with
fluctuations of about 0.04~MHz, though the standing waves only have
about 0.3\% in amplitude for the 19-beam system~\citep[][]{fast20}. In
this paper, we fit and subtract the band-pass profile with the
asymmetrically re-weighted penalized least squares smoothing (ArPLS)
algorithm \citep{arpls,zcl+21}, which was developed for the baseline
correction in the spectral analysis. Different approaches were tested
to deal with the complex band-pass profiles of the GPPS spectra
(e.g. higher-order polynomial, sinusoidal baseline-fitting), and the
ArPLS algorithm \citep{arpls,zcl+21} is found to be currently the most
optimal solution for a good baseline fitting and standing waves
removal \citep[see also][]{Hong2022}.

The RFI mitigation in the data processing includes three steps:
(1) discarding the RRL spectra suffering from severe RFI. The output
from the ArPLS fitting \citep{arpls,zcl+21} is the spectrum with
RFI. We first calculate the rms values for the spectra of
H165$\alpha$~$-$~167$\alpha$ RRLs which are almost free from strong
RFI. The line-free channels for calculating the spectral rms are
approximately between the V$_{\rm LSR}$ ranges of $-200$ to
$-160$~km~s$^{-1}$ and $160$ to 200~km~s$^{-1}$. Then, any RRL spectra
with a rms five times larger than the median rms of the
H165$\alpha$~$-$~167$\alpha$ are discarded since they are RFI
contaminated, especially H172$\alpha$~$-$~177$\alpha$ as shown in
Figure~\ref{lines}.
(2) removing RFI with fluctuations of about 0.04~MHz mainly for the
spectra recorded by beams No.~08 and No.~02 before March 02, 2022.
They result in a larger spectral rms value (typically around 0.4~mJy)
and influence the identifications of weak RRLs. This narrow RFI
signature can be easily identified from the discrete Fourier transform
of the observed spectra, then it will be removed and replaced by
interpolation. The data are then transformed back to form the
corrected spectra.
(3) removing the narrow-band RFI. For each of the RRL spectra, a
median filter with a kernel size of 9 channels is applied to a spectrum,
then the spectral channels with values greater than 5 times the
spectral rms relative to the median value are masked and replaced by
interpolation. In some difficult cases, a different method for removing the narrow-band RFI was
also applied to the spectra for a comparison: we fit and subtract the
band-pass profile with the ArPLS algorithm, and the spectral channels
with a value greater than 3.5 times of the rms level are discarded.

After these steps, most RFI has been removed although weak RFI signals
may still be left in some of the RRL spectra. The cleaned RRL spectra
are then used for stacking in the following steps.

\subsection{Baseline fitting and RRL stacking}

In the normal observations of spectral lines, an ``off-source''
reference spectrum is often needed to eliminate the complex band-pass
profile of the ``on-source'' spectrum in order to better identify the
weak signals of spectral lines.
Being different from the position-switching, beam-switching, or
``on-the-fly'' mapping modes commonly adopted in spectral line
observations, the GPPS survey does not have ``off-source'' integration
during the snapshot observations, which would be very time-consuming
otherwise. It is always difficult to find a reasonable clean
``off-source'' reference spectrum of a beam from the GPPS data to
perform the traditional ``on-off'' subtraction because of that: (1)
many of the RRL sources are much extended compared to the telescope
beam size; (2) the four sequential pointings of a certain beam in a
snapshot are adjacent to each other in their targeted positions in the
sky, looking like one ``on'' with three ``off'', but for such diffuse
emission of ionized gas, this does not help much; (3) the response and
band-pass properties of different beams have their own
characteristics, as shown in Figure~\ref{gain}, and therefore cannot
be used in mixture. In addition, the band-pass is also very diverse
for different RRL frequencies.

With the development of instruments in recent years, such an ``off''
reference spectrum may not be essential for spectral line
observations, because modern radio telescopes have good stability of
the band-pass. Instead of performing the traditional ``on-off''
subtraction, the narrow spectral lines can be reliably identified and
calibrated merely from the ``on-source'' integration spectrum alone
after a baseline fitting and removal \citep[e.g.,][]{pagani20}, so
that the observing efficiency can be significantly improved. This
approach has been tested and found to be applicable to the piggyback
spectral line data of the FAST GPPS survey.

The RRL stacking has been widely used in various studies to improve
the observation sensitivity and survey efficiency
\citep[e.g.,][]{bal06, hipass, thor, siggma, gdigs}.
Because the RRLs in the FAST observation band are very weak in nature,
it would be very time-consuming to accumulate only one RRL to a high
signal-to-noise ratio.
For the concerned $L$-band RRLs, H$165\alpha$ to H$183\alpha$, the RRL
intensity is expected to differ by less than 30\% without considering
the beam size effects \citep[Equation A7 of][]{gdigs}. As these RRLs
originate from the same astronomical sources, their line widths and
velocities should resemble each other. In the first order
approximation, it is reasonable to average all detectable RRLs to get
a mean spectrum, which can significantly improve the signal-to-noise
ratio.

In the practical data processing of the piggyback spectral lines of
the GPPS survey, two slightly different approaches have been tested.

In the first approach, the baseline of each usable RRL spectra with
the 5-minute integration of each beam was fitted and subtracted. Then the
RRL spectra are re-sampled with a velocity resolution of
2.2~km~s$^{-1}$, and added together to derive a mean spectrum. We
noticed that the averaged spectra sometimes show visible pits near the
line-wing regions caused by the overestimates of the baseline
level. Though the pit is almost negligible for a single RRL spectrum,
but it becomes significant after adding all RRL spectra. Therefore
this approach is not optimal for processing the RRL data.

A modified method is developed. We make data stacking for these RRLs
first and then do the baseline fitting and removal. The modified
method consists of four steps: (1) re-sampling all usable RRL spectra
with the same velocity resolution of 2.2~km~s$^{-1}$ which is the
value for H$183\alpha$; (2) adding all re-sampled spectra together to
generate a combined spectrum, with the weighting factor of integration
time; (3) fitting and subtracting the baseline with the ArPLS
algorithm \citep{arpls,zcl+21}, so that the broader standing waves
($\sim$1~MHz) mentioned in \citet{fast20} can be naturally removed in
this step; (4) deriving a mean spectrum of RRLs. In this step, the
baselines around each RRL are not modified. The baseline of the
combined spectrum has a much higher signal-to-noise and can be fitted
and removed more easily. In addition, the systematic dip occasionally
seen with the first method disappears.

The modified method is also applied separately to the first
eight (H$165\alpha$ to H$172\alpha$) and the remaining 11 RRL
spectra (H$173\alpha$ to H$183\alpha$), to derive two sets of averaged
RRL spectra. If the signatures of two spectra are consistent, they are
taken as a reliable detection. Otherwise, the inconsistency must be
caused by the residual RFI after the mitigation and all RRL data must
be re-processed.

\begin{sidewaysfigure*}
  \centering
  \vspace{-9.8cm}
  \includegraphics[width=200mm]{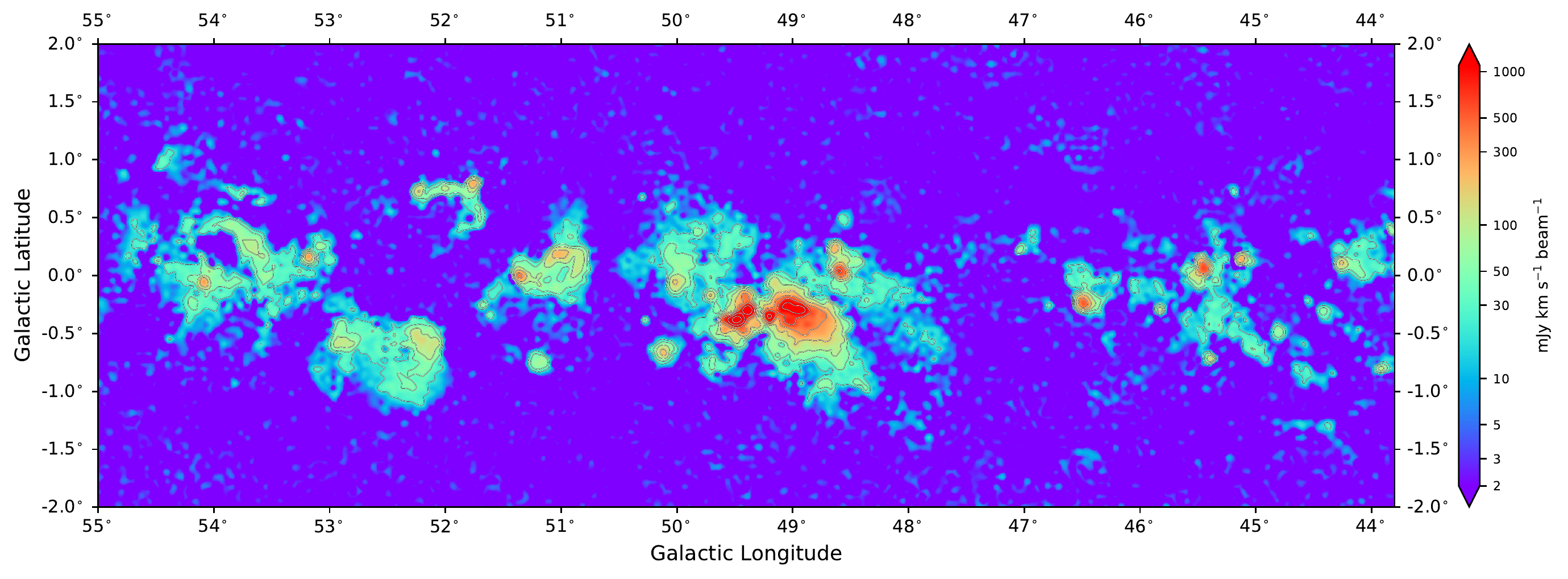}\\
  \vspace{-0.18cm}
   \includegraphics[width=200mm]{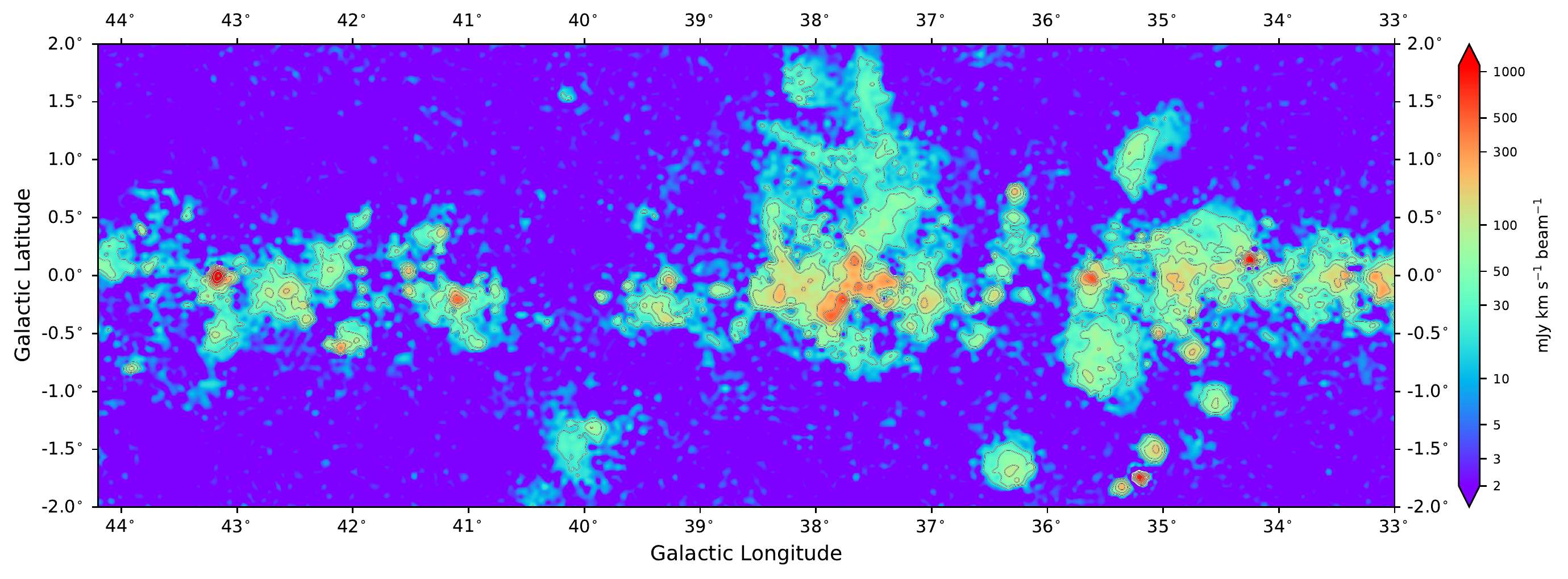}\\  
   \caption{Velocity-integrated intensity maps of the H$n\alpha$ RRLs
     recorded by the FAST GPPS survey \citep[][]{gpps} for an inner
     Galaxy region of $33^\circ \leqslant l \leqslant 55^\circ$, $|b|
     \leqslant 2\fdg0$. { The observation beam has a size of about
       3$^\prime$}. The velocity range for the integration is from
     $V_{\rm LSR} =$~$-40$~km~s$^{-1}$ to 120~km~s$^{-1}$. The
     overlaid contour levels are at
     2$^n\times5$~mJy~km~s$^{-1}$~beam$^{-1}$, with $n=2,3,...10$. }
   \label{intall}
\end{sidewaysfigure*}

\begin{figure*}
  \centering
  \includegraphics[width=0.74\textwidth]{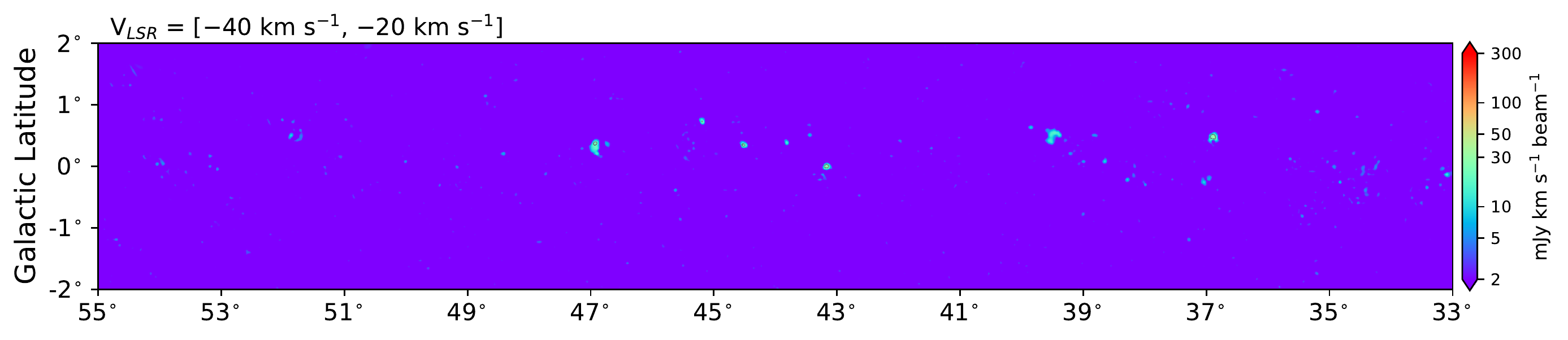}\\
  \vspace{-1.85mm}
  \includegraphics[width=0.74\textwidth]{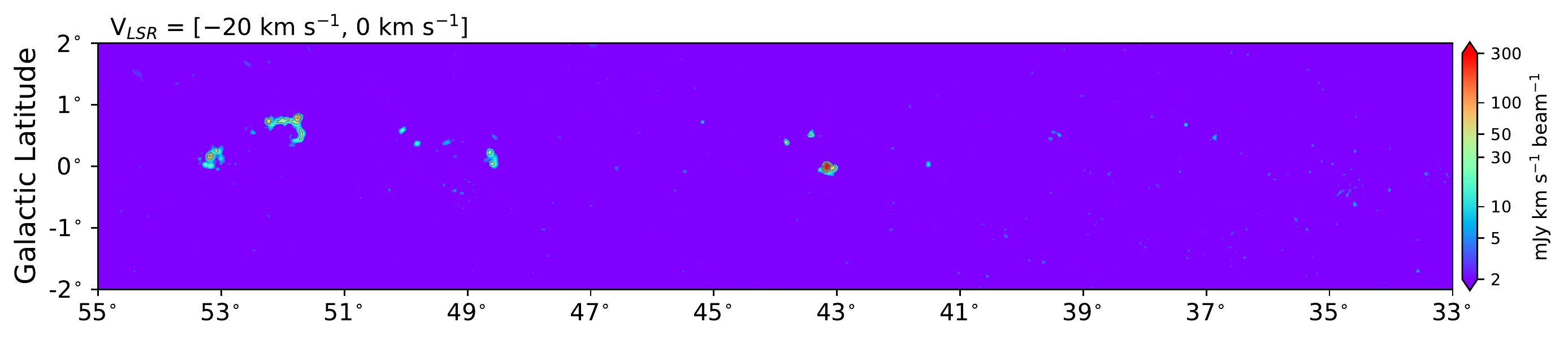}\\
  \vspace{-1.85mm}
  \includegraphics[width=0.74\textwidth]{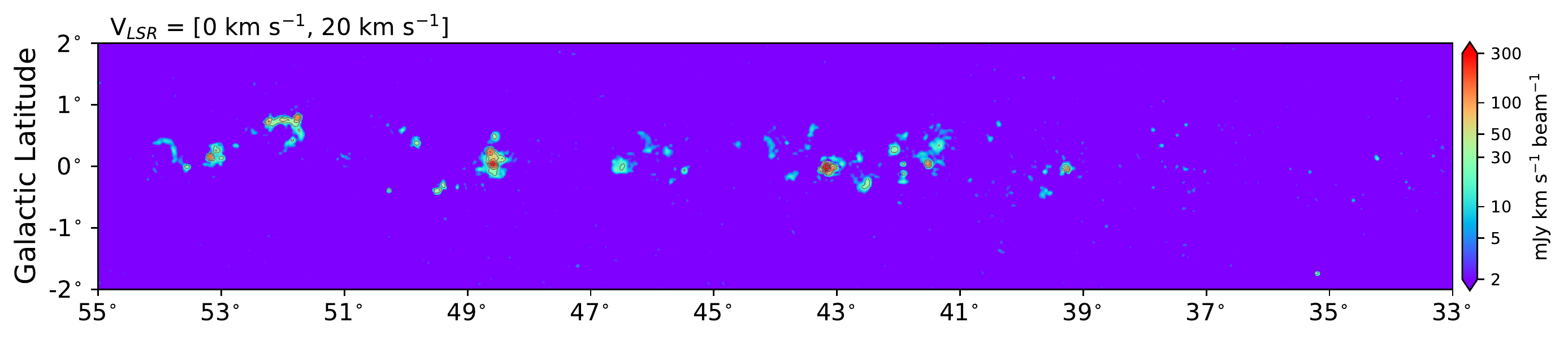}\\
  \vspace{-1.85mm}
  \includegraphics[width=0.74\textwidth]{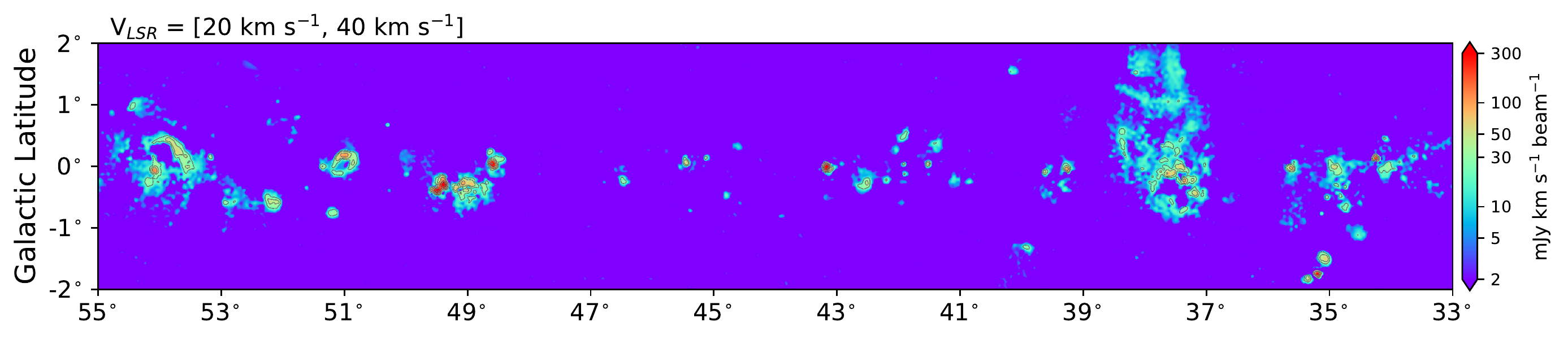}\\
  \vspace{-1.85mm}
  \includegraphics[width=0.74\textwidth]{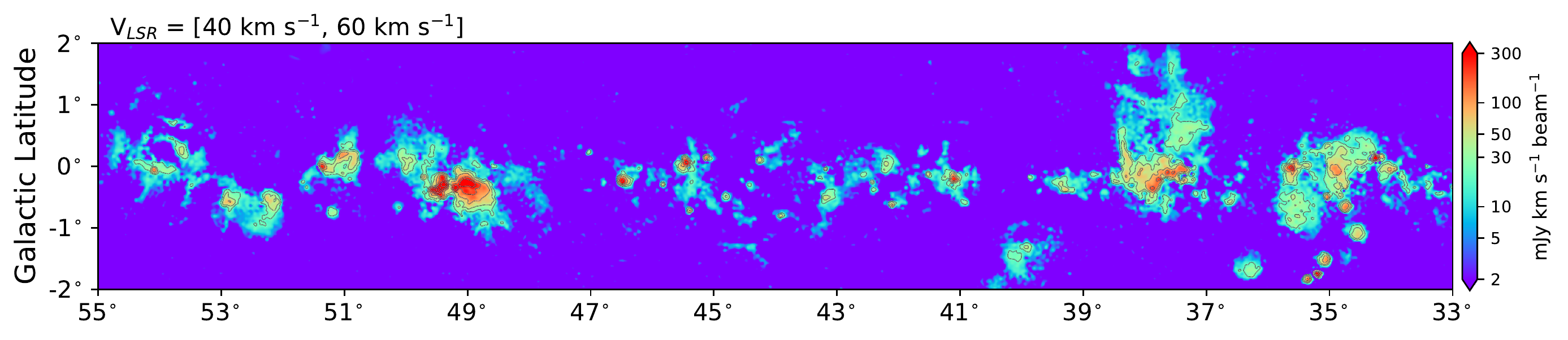}\\
  \vspace{-1.85mm}
  \includegraphics[width=0.74\textwidth]{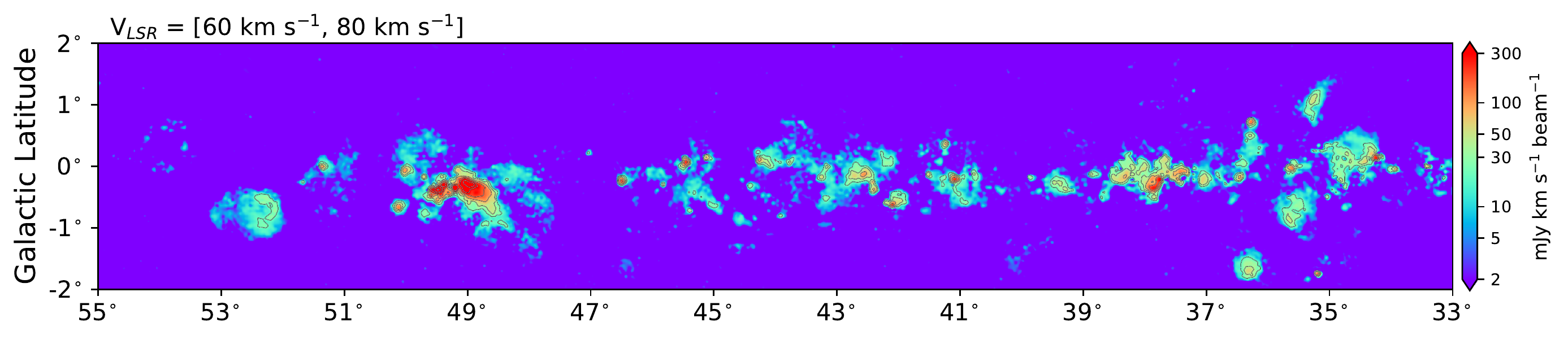}\\
  \vspace{-1.85mm}
  \includegraphics[width=0.74\textwidth]{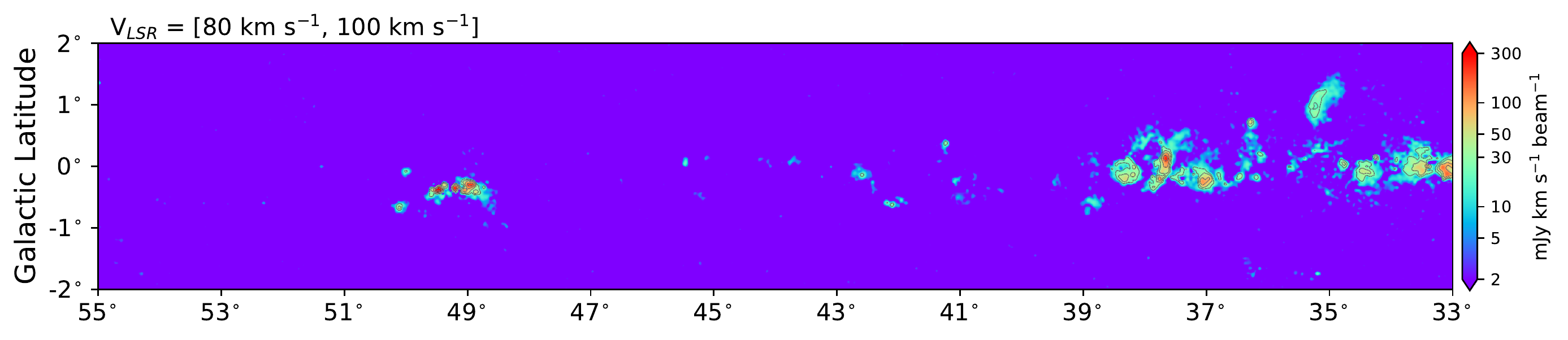}\\
  \vspace{-1.85mm}
  \includegraphics[width=0.74\textwidth]{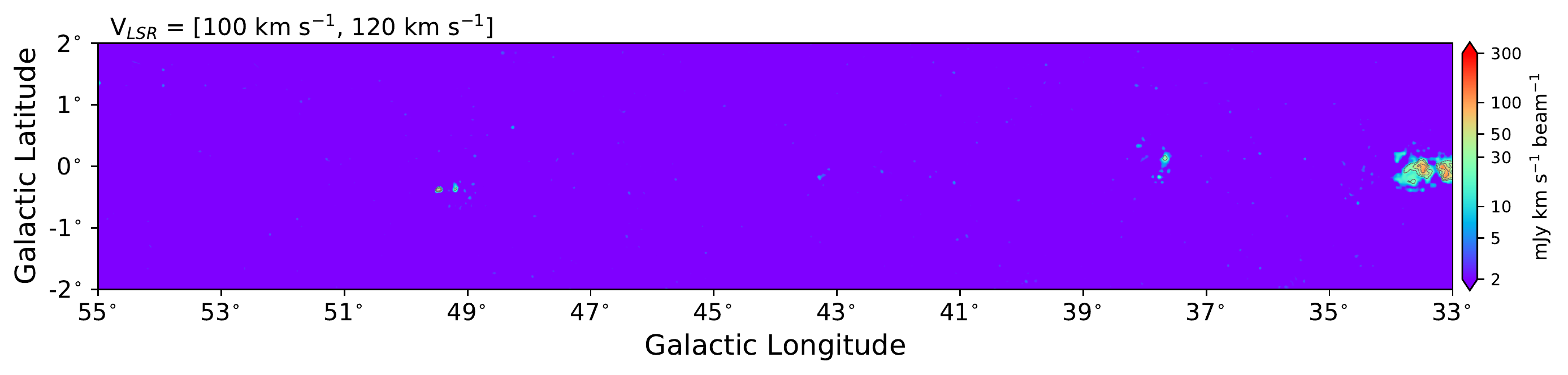}\\
  \caption{Channel maps of the H$n\alpha$ RRLs for an inner Galaxy
    region of $33^\circ \leqslant l \leqslant 55^\circ$, $|b|
    \leqslant 2\fdg0$, integrated over a channel width of
    20~km~s$^{-1}$ in the velocity range marked on the top of each
    panel. The contour levels are the same as that of
    Figure~\ref{intall}. The FAST observation beam has a size of about
    3$^\prime$.}
   \label{channelmap}
\end{figure*}

\begin{figure*}[t]
  \centering
   \includegraphics[width=0.98\textwidth]{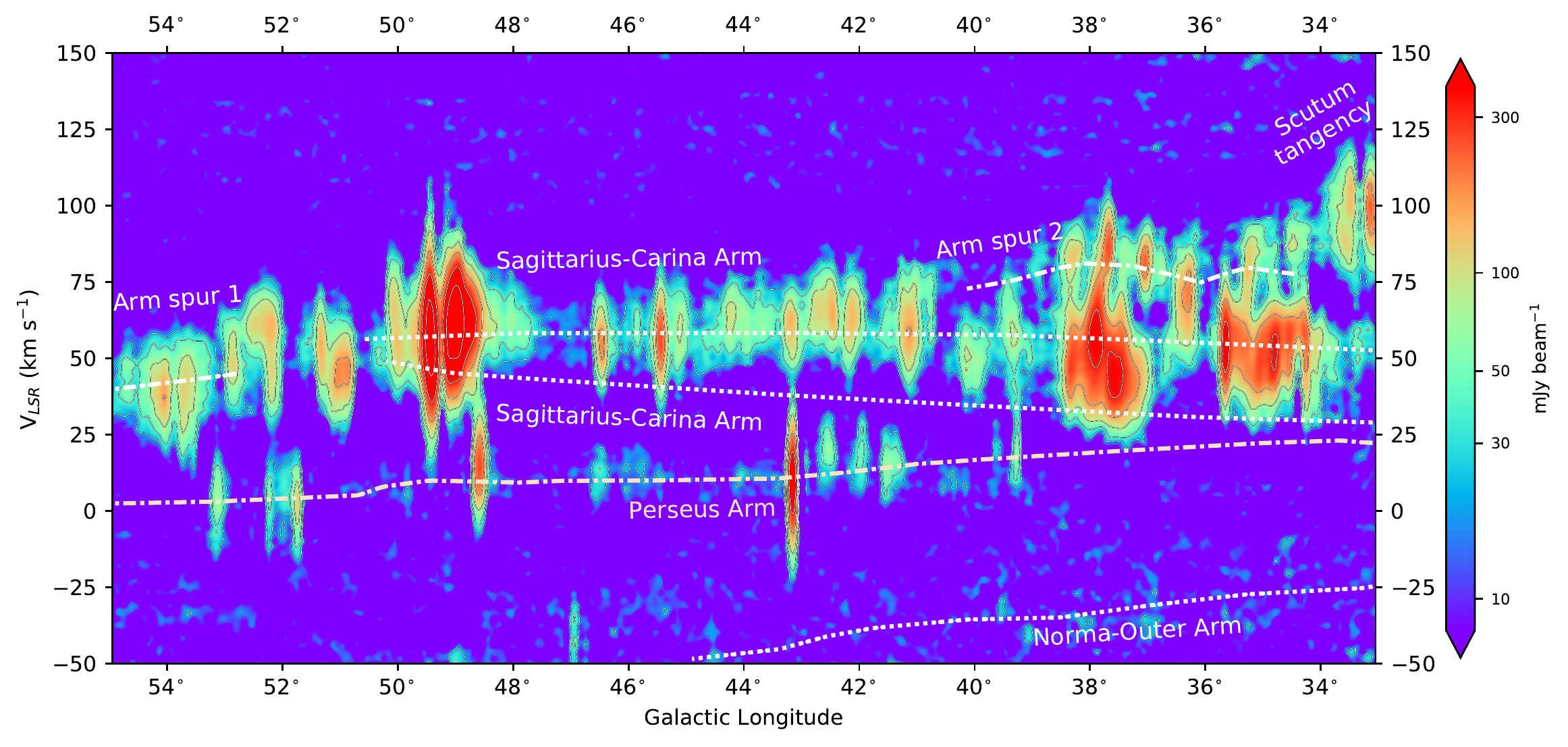}\\
   \caption{Longitude–velocity diagram of the H$n\alpha$ RRLs recorded
     by the FAST GPPS survey \citep[][]{gpps} for an inner Galaxy
     region of $33^\circ \leqslant l \leqslant 55^\circ$, $|b|
     \leqslant 2\fdg0$. A spiral arm model given by \citet{rmb19} is
     outlined: the Sagittarius-Carina Arm (dotted lines), the Perseus
     Arm (dash-dotted line), the Norma-Outer Arm (dotted line), and
     two arm spurs (dash-dotted lines). One spur is located between
     the Perseus Arm and the Sagittarius Arm (arm spur 1), the other
     is between the Sagittarius Arm and the Scutum Arm (arm spur 2). A
     part of the Scutum tangency region \citep[e.g.,][]{hh15} covered
     by the RRL data is also marked. The overlaid contour levels are
     2$^n\times5$~mJy~beam$^{-1}$ with $n=2,3,...7$. The features with
     $V_{\rm LSR} < -50$~km~s$^{-1}$ are dominated by He$n\alpha$ and
     C$n\alpha$ RRLs, hence are not presented here.  }
   \label{lvmap}
\end{figure*}

\subsection{Calibration and systematical uncertainties}

In the calibration of the intensity scale from $T_a$ to Jy with 
respect to the standard flux calibrators of 3C\,286 and 3C\,138, we
adopted the averaged gain curves from the real FAST observations. The
relative uncertainties in principle are small, while we have to
estimate the uncertainty of absolute scales.

Two methods of calibration have been involved in data
processing. The first is using the periodically injected reference 
signal to convert the XX and YY polarization products to the intensity
units of $T_a$. The noise diode in the signal path of the FAST 19-beam
system has temperature fluctuations on the order of $<$1\% over
several hours \citep{fast20}. The gain fluctuations of the $L$-band 19-beam system are
typically on the order of a few percent over a timescale of a few
minutes and less than about 4\% over several hours \citep{fast20}.
The uncertainties induced by the injection of the reference signal as
well as the long term fluctuations of the system gain curves, could
cause an overall uncertainty for the absolute RRL intensity of less than
$\sim$10\%.

The $\hi$ data recorded by the FAST GPPS survey have also been
processed by \citet{Hong2022}. The $\hi$ results of the GPPS survey
have been compared with those of the Effelsberg-Bonn $\hi$ Survey
\citep[][]{ebhis16}. Despite the not-yet-corrected stray radiation,
the systematical differences in intensity scales between different
FAST snapshot observations and between different beams of the 19-beam
receivers have been noticed. It is probably caused by different
characteristics of the long term fluctuations of the system gain
curves for different beams.  The comparison of the FAST GPPS $\hi$
results with those of the Effelsberg-Bonn $\hi$ Survey
\citep[][]{ebhis16} gives the correction factors for different beams
in different covers. After the corresponding scale factors are
applied, the final results are compared again and found to be
consistent within 3\%. These correction factors are also adopted in
this work to the data products of H$n\alpha$ RRLs.

\section{Results}
\label{results_sec}

The GPPS survey piggyback line data in the area of Galactic longitude
$33^\circ \leqslant l \leqslant 55^\circ$ and latitude $|b| \leqslant
2\fdg0$ have been processed. More than { 43\,000} averaged spectra of
H$n\alpha$ RRLs are obtained. The derived RRL spectra have been
visually checked one by one to ensure the correctness of data
processing. The velocity-integrated intensity map is shown in
Figure~\ref{intall}, which is summed from $V_{\rm LSR} =
-40$~km~s$^{-1}$ to 120~km~s$^{-1}$ of the H$n\alpha$ RRL
spectra. More detailed channel maps are shown in
Figure~\ref{channelmap}.

\begin{figure}[H]
  \centering \includegraphics[width=0.47\textwidth]{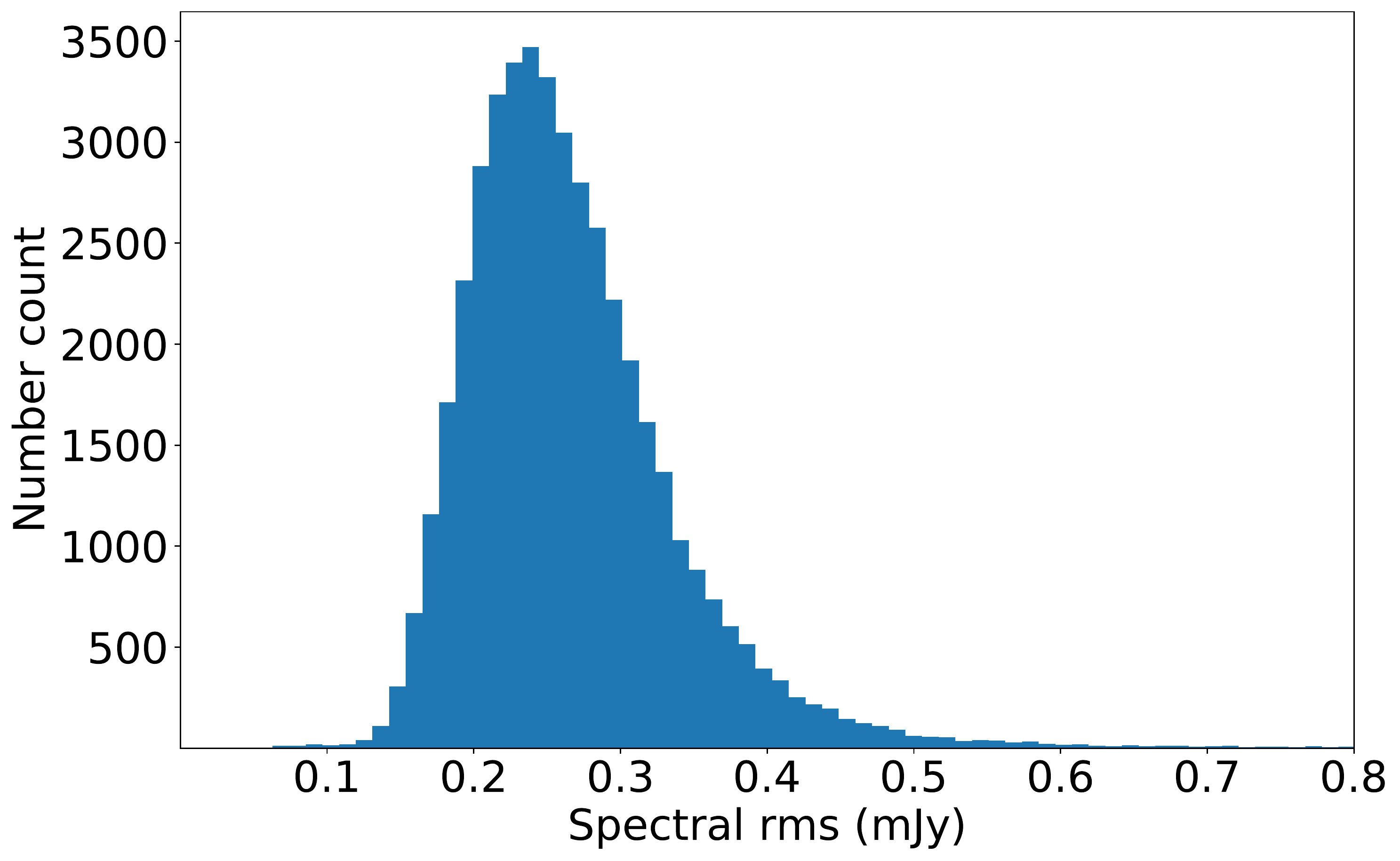}
  \caption{Distribution of the spectral rms values calculated from the
    line-free channels approximately between the $V_{\rm LSR}$ ranges
    of $-200$ to $-160$~km~s$^{-1}$ and $160$ to 200~km~s$^{-1}$ with
    a spectral resolution of 2.2~km~s$^{-1}$, for all beams in the
    mapped area of $33^\circ \leqslant l \leqslant 55^\circ$ and $|b|
    \leqslant 2\fdg0$. The typical value is about
      0.25~mJy~beam$^{-1}$. The tail in the higher end is
    primarily caused by bright continuum radio sources.}
   \label{rms}
\end{figure}

In the map there are complex structure features dominated by a number
of $\hii$ regions as well as diffuse ionized gas. 
Several well-known and large star-forming complexes are prominent in
the map, for example, W\,51 around ($l,b$)~$\sim$~($49\fdg1$, $-0\fdg6$),
W\,49 around ($l,b$)~$\sim$~($43\fdg2$, $-0\fdg0$) and
W\,47 around ($l,b$)~$\sim$~($37\fdg8$, $-0\fdg2$).
Beside the discrete $\hii$ regions, the diffuse component of ionized
gas are widely distributed, for example, towards a higher latitude
region of $b > 2^\circ$ in the Galactic longitudes around
$l\sim38\fdg0$.

The global distribution and the dynamics of the Galactic ionized gas
associated with star formation regions can be used to trace the spiral
structure. The longitude–velocity diagram of the RRL data is shown in
Figure~\ref{lvmap}, with indications of spiral arms in the model of
\citet{rmb19}.
The detected RRL features are primarily related to the
Sagittarius-Carina Arm, the Perseus Arm, as well as two arm spurs. One
arm spur is located between the Perseus Arm and the Sagittarius Arm,
and the other is located between the Sagittarius Arm and the Scutum
Arm. The RRL data also cover part of the Scutum tangency region.
Some RRL clouds possess a negative velocity, indicating that they are
located at very distant spiral arms in the outer Galaxy, i.e. with a
Galactocentric distance larger than the solar orbit.

In the map shown in Figure~\ref{intall}, the noise level is generally
uniform, except for some regions associated with relatively strong
continuum emission in the bright $\hii$ regions of G34.256+0.136 and
G45.453+0.045, in the star-forming complexes of W\,51 and W\,49, and
in supernova remnants of W\,44 around ($l,b$)~$\sim$~($34\fdg7$, $-0\fdg4$),
W\,49b around ($l,b$)~$\sim$~($43\fdg3$, $-0\fdg2$), G39.2$-$00.3 (or 3C\,396) and G49.2$-$0.7.

The rms values for the RRL spectra are calculated from the line-free
channels approximately between the $V_{\rm LSR}$ ranges of $-200$ to
$-160$~km~s$^{-1}$ and $160$ to 200~km~s$^{-1}$ with a spectral
resolution of 2.2~km~s$^{-1}$.
The distribution of spectral rms values from all beams in the
processed sky area is shown in Figure~\ref{rms}. The extension of the
peak is partially caused by the RFI affected channels or other factors
such as the stability of the focus cabin. The tail in the higher end
is primarily caused by bright continuum radio sources. The typical
spectral rms is about 0.25~mJy~beam$^{-1}$, estimated by a Gaussian
fitting to the rms distribution. The rms value can be compared with
other RRL surveys at $L$ band, which is about 26\% of the rms level of
the Arecibo SIGGMA survey \citep[][]{siggma}, about 4.1\% of that of
the VLA THOR survey \citep[][]{thor}, and about 1.3\% of the Parkes
HIPASS survey \citep[][]{hipass}, if these values are calculated in
the same spectral resolution of 2.2~km~s$^{-1}$ (see
Figure~\ref{comsen}). In other words, the FAST RRL observation
sensitivity is about 4 times better than the SIGGMA survey, and 20
times better than the THOR survey.

\begin{figure}[H]
  \centering
\includegraphics[width=0.43\textwidth]{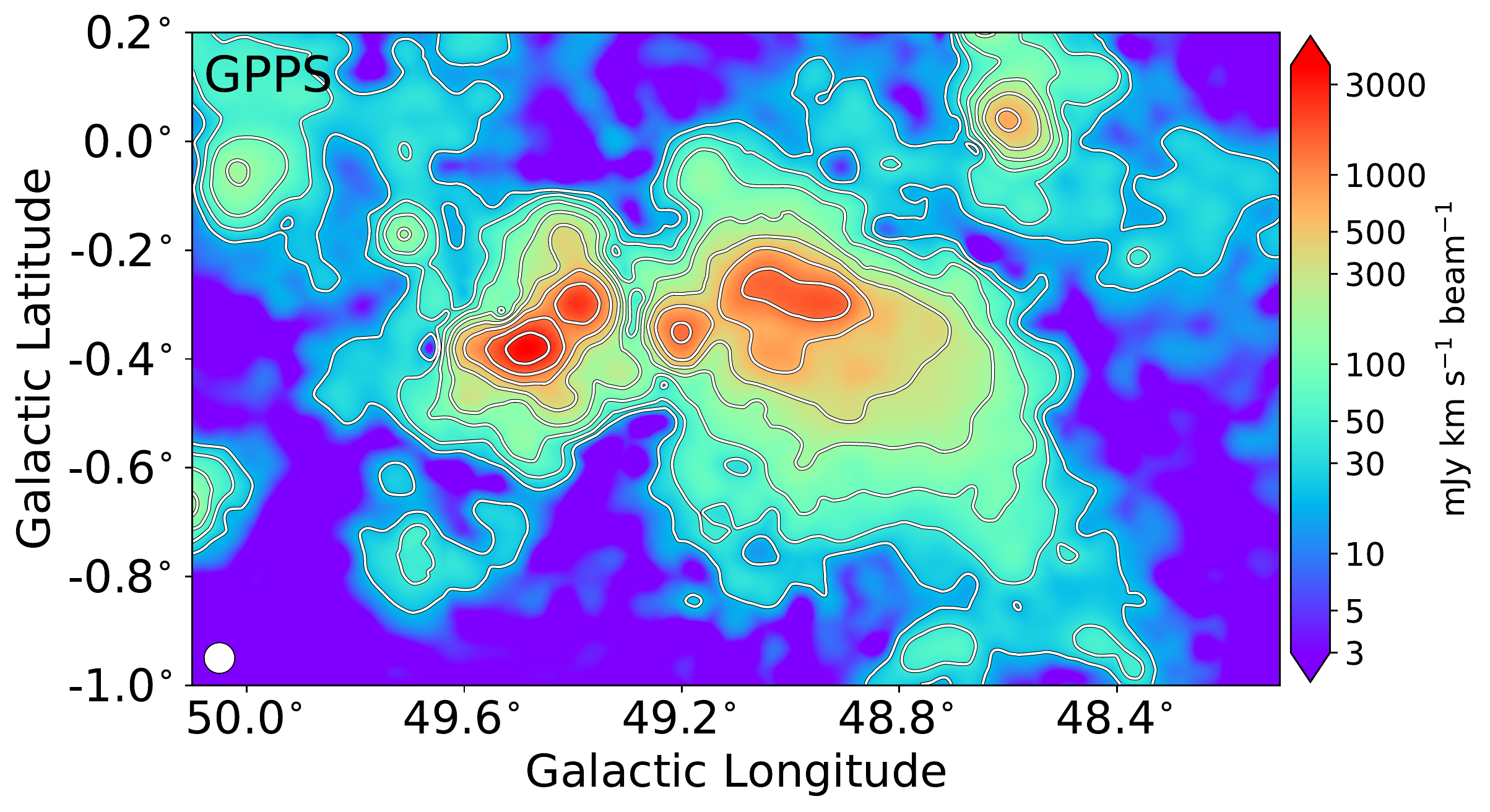}\\
\vspace{-0.39cm}
\includegraphics[width=0.43\textwidth]{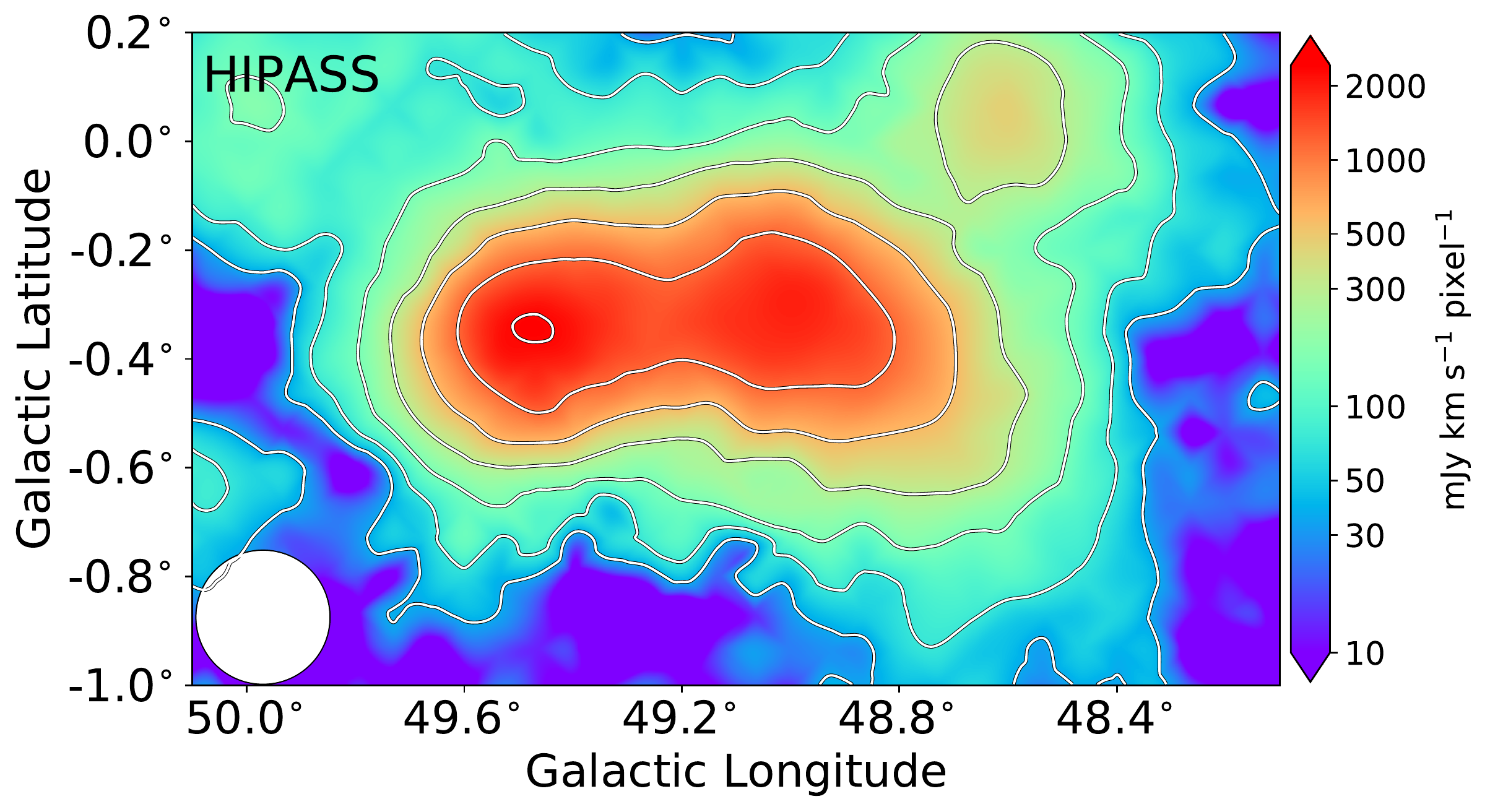}\\
\vspace{-0.39cm}
\includegraphics[width=0.43\textwidth]{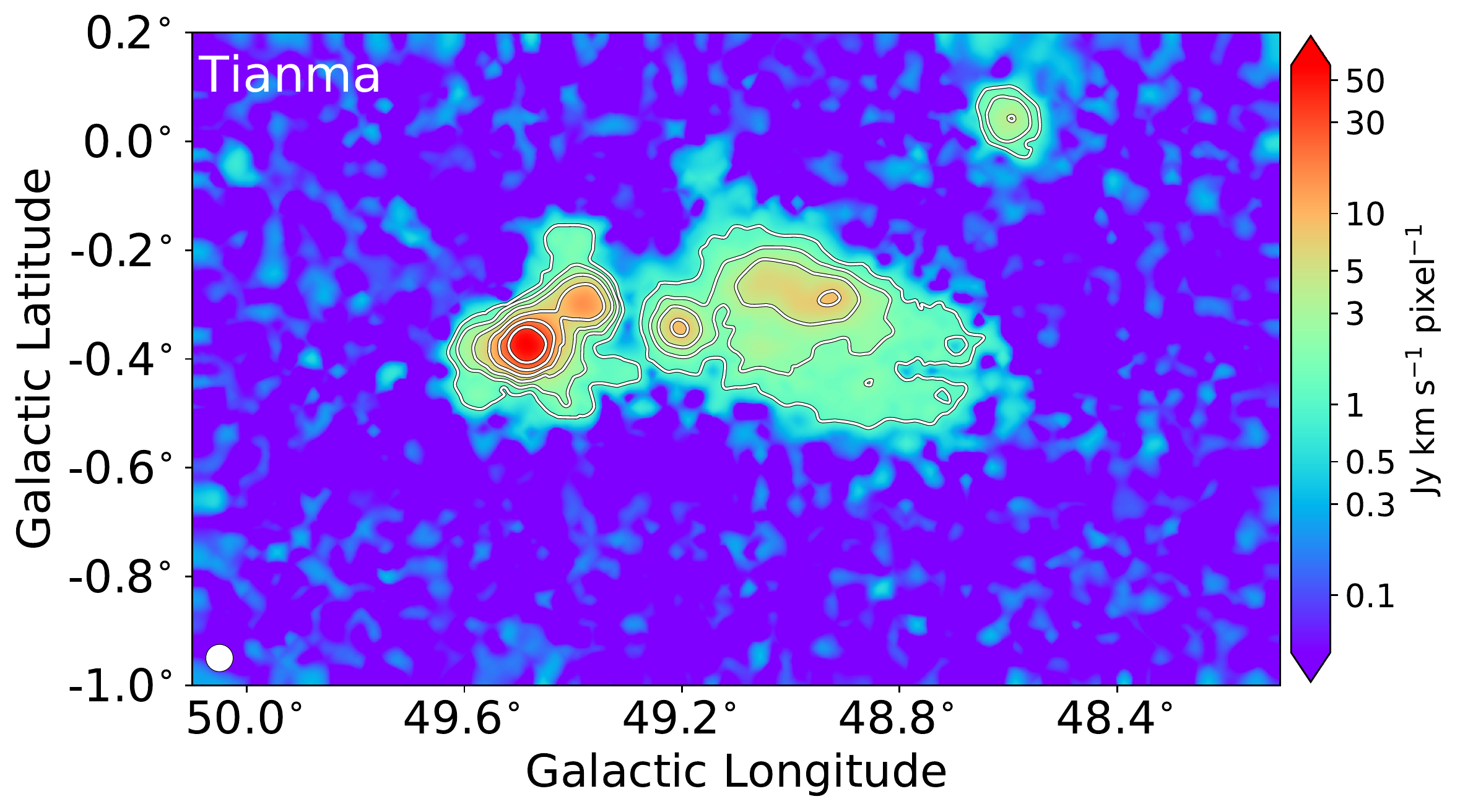}
   \caption{ Velocity-integrated intensity maps of RRLs for a
     star-forming complex, W\,51, obtained by the FAST GPPS survey at
     $L$ band ({\it upper panel}), compared to the maps obtained by
     the HIPASS project at $L$ band~\citep[{\it middle
         panel,}][]{hipass}, and the Tianma 65-m radio telescope at
     $C$ band~\citep[{\it lower panel,}][]{hou17}. The overlaid
     contour levels are at 2$^n\times5$~mJy~km~s$^{-1}$~beam$^{-1}$
     ($n=2,3,...10$), 2$^n\times4.5$~mJy~km~s$^{-1}$~pixel$^{-1}$
     ($n=3,...10$), and 2$^n$~Jy~km~s$^{-1}$~pixel$^{-1}$
     ($n=0,1,2,3,4,5,6$) for the {\it upper, middle} and {\it lower}
     panels, respectively. The telescope beam size is indicated in the
     lower-left corner of each plot with a white filled circle.}
\label{w51com}
\end{figure}

The sensitivity of the FAST GPPS RRL survey can also be expressed by
the main-beam brightness temperature $T_B$ and the emission measure
(EM). The brightness temperature $T_B$ is related to the flux density
via $T_B=({\lambda^2S_\nu}) / ({2k_B\Omega_{bm}})$. { Here, $\lambda$
  is the wavelength at an averaged frequency of about 1.24~GHz for the
  FAST RRLs}, $S_\nu$ is the flux density, $k_B$ is the Boltzmann
constant, and $\Omega_{bm}$ is the beam solid angle. By assuming a
Gaussian beam, the relation can be simplified as ${T_B}/{K} \approx
1.36 ({\lambda}/{cm})^2 ({S_\nu}/{\rm mJy}) / ({\theta^2})$, { where
  $\theta \approx 180''$ is the half-power beam width for the FAST RRL
  observations}. We get $T_B = 6.3$~mK for the main-beam brightness
temperature for the GPPS H$n\alpha$ RRL sensitivity, as shown in
Figure~\ref{comsen}. { The brightness temperature sensitivity for
  other RRL surveys are estimated through the same method with the
  adopted values of mean observation frequency and $\theta$ as being
  1.40~GHz and $864''$ for HIPASS, 1.37~GHz and $360''$ for SIGGMA,
  5.76~GHz and $159''$ for GDIGS, and 5.92~GHz and $180''$ for
  GLOSTAR-Eff, respectively.}
To estimate the physical units of emission measure, EM~=~$\int n_e^2
dl$, for the RRL surveys with single-dish telescopes, we follow the
equations (A4) to (A7) given in the appendix of \citet{gdigs} and the
procedures described in their Sect.2.6. Here, $n_e$ is the electron
density, $l$ is the path length. The plot for $T_B$ versus EM of all
surveys is given in Figure~\ref{comsen}, including the GBT GDIGS
project at $C$ band \citep[][]{gdigs} which has the lowest spectral
rms in $T_B$ ($\sim$ 3.6~mK) among the discussed RRL surveys. The RRL
map from the piggyback data of the FAST GPPS project is currently the
most sensitive at $L$ band, down to an emission measure of about
$200$~cm$^{-6}$~pc. Therefore it has the advantage to detect very low
electron density regions, see an example region in
Figure~\ref{w51com}.

In the following, we discuss some details of the Galactic $\hii$
regions and the distribution of diffuse ionized gas.

\begin{figure*}[t]
  \centering
  \includegraphics[width=0.92\textwidth]{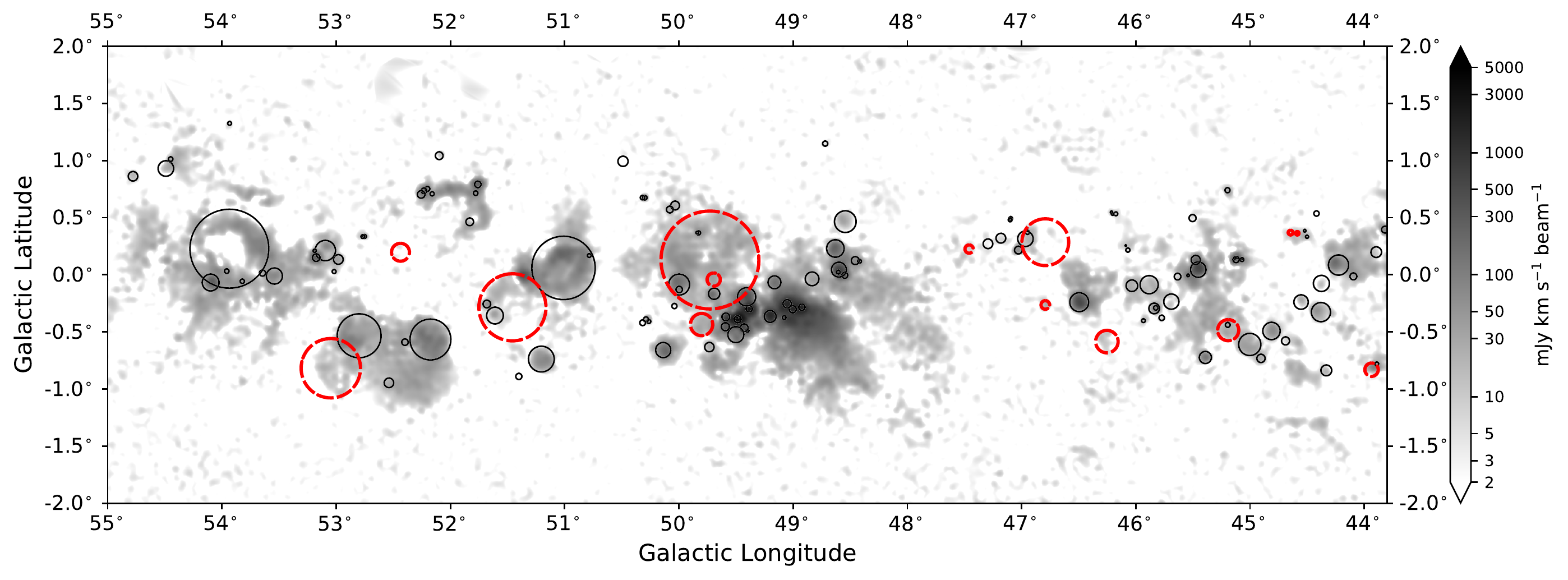}\\
  \vspace{-5mm}
  \includegraphics[width=0.92\textwidth]{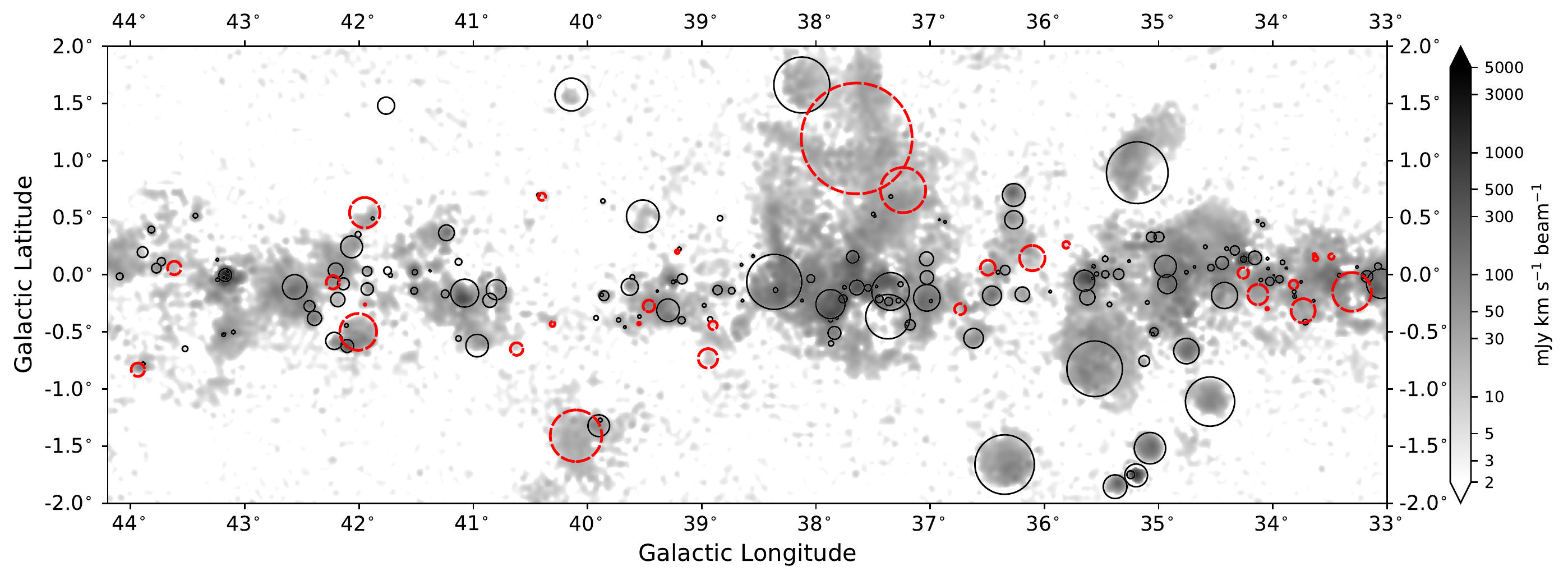}  
  \caption{Distributions of the known $\hii$ regions (solid line
    circles, Sect.~\ref{known}) and newly confirmed $\hii$ regions
    (red dashed circles, see Sect.~\ref{new}) marked on the averaged
    RRL image from the piggyback spectral data of the GPPS project in
    the Galactic area of $33^\circ \leqslant l \leqslant 55^\circ$ {
      and $|b| \leqslant 2\fdg0$}. The circle size stands for the
    $\hii$ region radius given by the {\it WISE} catalogue of $\hii$
    regions and candidates \citep[][]{wise14}. { The background image
      is a greyscale version of} Figure~\ref{intall}. }
   \label{hii-distribution}
\end{figure*}

\begin{table*}[!t]
\centering
\renewcommand\arraystretch{0.93}
\footnotesize
\caption{H$n\alpha$ RRL parameters for 302 known $\hii$ regions FAST-GPPS detected in the Galaxy area of $33^\circ \leqslant l \leqslant   55^\circ$ and $|b| \leqslant 2\fdg0$.}
\label{knownhii}
\vspace{1mm}
\tabcolsep8pt 
\begin{tabular*}{\textwidth}{ccrrrcccccr}
\toprule
\hline 
$\hii$ region & $l$          & \multicolumn{1}{c}{$b$}        & Radius  & V$_{\rm LSR}^{\rm Ref.}$  &  Ref. &  Peak intensity    & V$_{\rm LSR}$    & FWHM         &  rms      & S/N  \\
              & $(^\circ)$    & \multicolumn{1}{c}{$(^\circ)$}   & ($^{\prime\prime}$)  &  (km~s$^{-1}$)  &  & (mJy~beam$^{-1}$)  & (km~s$^{-1}$)  & (km~s$^{-1}$)  & (mJy~beam$^{-1}$)  &  \\
(1)           &   (2)  &     (3)   &  (4) &  (5) &                 (6)   &          (7)    &         (8)       &         (9)     &  (10) & (11)  \\
\hline
G33.051$-$0.078 & 33.051 &  $-$0.078 &  466 &  93.8&      8   &   6.8 $\pm$ 0.2 &   95.3 $\pm$ 0.3  &  26.3 $\pm$ 0.8 &  0.43& 15.9\\
G33.080+0.073   & 33.080 &     0.073 &  110 &  86.9& 1 &   7.7 $\pm$ 0.1 &   90.8 $\pm$ 0.3  &  28.0 $\pm$ 0.6 &  0.37& 20.9\\
G33.176$-$0.015 & 33.176 &  $-$0.015 &  185 & 105.8&  12 &  20.1 $\pm$ 0.2 &  100.2 $\pm$ 0.2  &  28.6 $\pm$ 0.4 &  0.59& 34.0\\
G33.205$-$0.012 & 33.205 &  $-$0.012 &   64 & 105.8&  12 &  14.3 $\pm$ 0.2 &  101.1 $\pm$ 0.2  &  27.3 $\pm$ 0.4 &  0.46& 31.1\\
G33.263+0.067   & 33.263 &     0.067 &   42 &  98.3& 1  &   2.0 $\pm$ 0.1 &  103.5 $\pm$ 0.8  &  23.8 $\pm$ 1.8 &  0.31&  6.5\\
G33.419$-$0.004 & 33.419 &  $-$0.004 &   60 &  76.5&      8   &   8.9 $\pm$ 0.2 &   73.6 $\pm$ 0.9  &  36.3 $\pm$ 1.9 &  0.56& 16.1\\ 
G33.643$-$0.228 & 33.643 &  $-$0.228 &   42 & 102.9& 1 &   2.9 $\pm$ 0.1 &  104.1 $\pm$ 0.5  &  21.1 $\pm$ 1.2 &  0.31&  9.5\\
G33.715$-$0.415 & 33.715 &  $-$0.415 &   89 &  53.3&  1 &   3.5 $\pm$ 0.2 &   53.0 $\pm$ 0.5  &  17.5 $\pm$ 1.1 &  0.37&  9.4\\
G33.753$-$0.063 & 33.753 &  $-$0.063 &   42 & 101.7& 1 &   5.3 $\pm$ 0.1 &  100.2 $\pm$ 0.3  &  22.5 $\pm$ 0.7 &  0.32& 16.5\\
...             &   ...  &     ...   &  ... &  ... &                 ...   &          ...    &         ...       &         ...     &  ... & ...  \\
\hline
\bottomrule
\end{tabular*}
\begin{tablenotes}
\item[1] { Notes.} The entire version in a machine-readable format is
  available at the webpage: {\it
    \color{blue}http://zmtt.bao.ac.cn/MilkyWayFAST/}. Column 1 is the
  source name; Cols. 2, 3 and 4 are the Galactic longitude, Galactic
  latitude and radius given in the {\it WISE} catalogue of $\hii$
  regions and candidates \citep[][]{wise14}; Col. 5 is the V$_{\rm
    LSR}$ velocity measured in the reference given in Col. 6. Ref.1 =
  \cite{hrds11}; Ref.2 = \cite{hrds2015a}; Ref.3 = \cite{hrds2015b};
  Ref.4 = \cite{hrds2017}; Ref.5 = \cite{araya02}; Ref.6 =
  \cite{balser11}; Ref.7 = \cite{arecibohii12}; Ref.8 = \cite{lock89};
  Ref.9 = \cite{lock96}; Ref.10 = \cite{qrbb06}; Ref.11 =
  \cite{sewilo2004}; Ref.12 = \cite{was03}; Cols. 7, 8, and 9 are the
  fitted peak intensity, V$_{\rm LSR}$ velocity and line width
  (full-width at half-maximum) by this work, together with their
  1$\sigma$ uncertainties; Col. 10 is the rms value of the averaged
  RRL spectrum; Col. 11 is the signal-to-noise ratio of the spectrum.
\end{tablenotes}
\end{table*}

\begin{figure*}[t]
  \centering
  \includegraphics[width=0.8\textwidth]{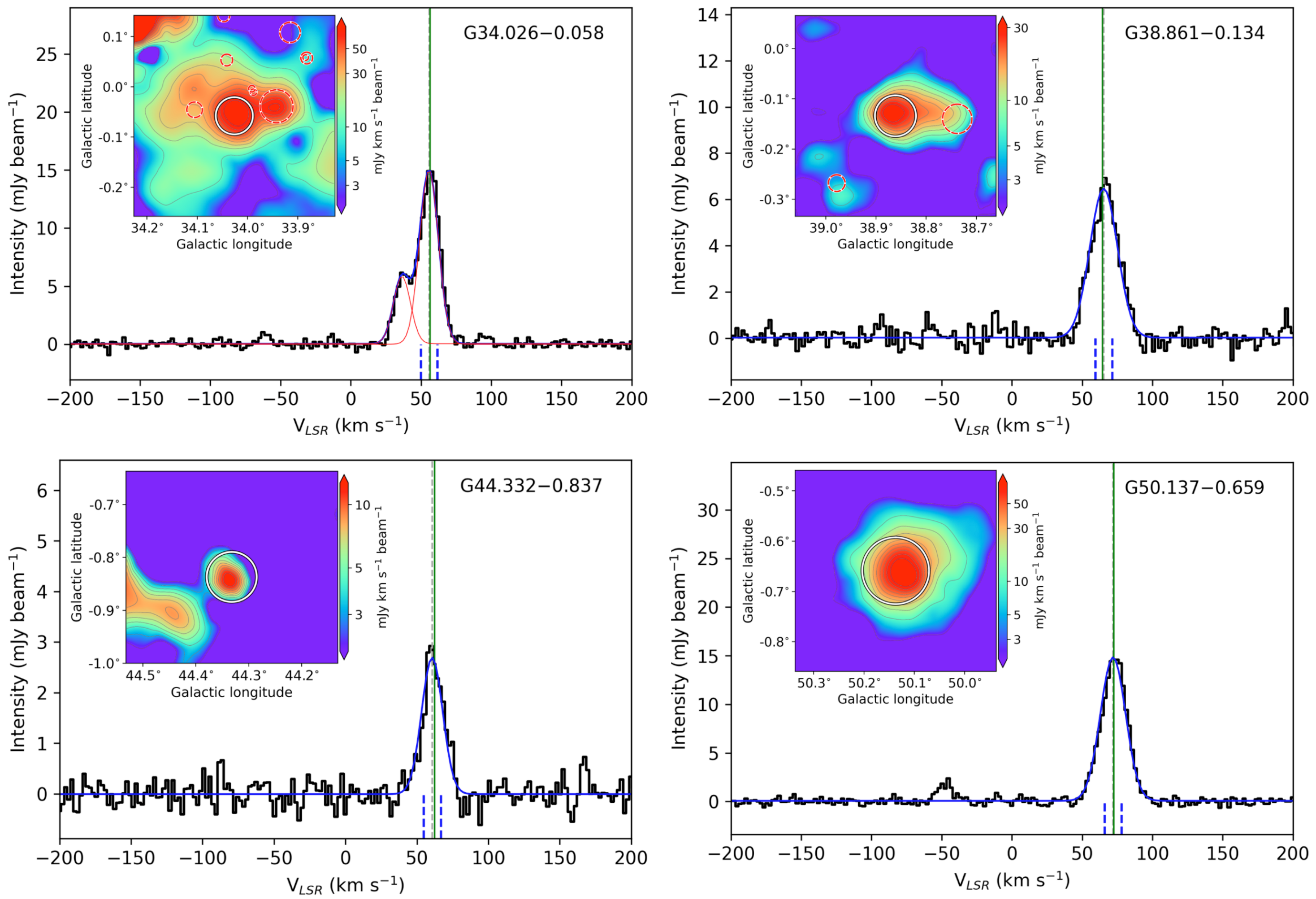}
  \caption{Examples of the RRL spectra towards known $\hii$ regions
    obtained by the piggyback spectral data of the FAST GPPS
    survey~\citep{gpps}. The line profile is fitted by Gaussian(s),
    and the V$_{\rm LSR}$ given in the references are indicated by a
    long solid vertical line and the V$_{\rm LSR}$ obtained from this
    work by the dotted line, which are almost overlapped. The two
    shorter dashed lines indicate the velocity range of V$_{\rm LSR}
    \pm 6$~km~s$^{-1}$ used to create the inserted velocity-integrated
    intensity map for known $\hii$ regions and the nearby area. In the
    intensity map the concerned $\hii$ region is shown by a solid line
    circle in the center. Other nearby known $\hii$ regions are
    indicated by dashed circles. The circle sizes are the source radii
    given in the {\it WISE} catalog \citep[][]{wise14}.}
   \label{hiiexample}
\end{figure*}

\subsection{RRL emission from known $\hii$ regions}
\label{known}

Up to now, more than 2\,000 Galactic $\hii$ regions have RRLs detected
\citep[e.g.,][]{dwbw80,wwb83, ch87,lock89,lock96,hrds,arecibohii12,
  hrds18,shrds19, shrds21, arm21}.
In the area of $33^\circ \leqslant l \leqslant 55^\circ$ { and $|b|
\leqslant 2\fdg0$}, the properties of Galactic $\hii$ regions have
been compiled by \citet{pbd03}, \citet{hh14}, \citet{wise14} and
\citet[][]{gao19}.
Among them, the {\it WISE} ({\it Wide-Field Infrared Survey Explorer})
catalog \citep{wise14} provides the largest data set of the Galactic
$\hii$ regions and candidates because all $\hii$ regions exhibit a
good mid-infrared morphology: the 12$\mu$m emission of an $\hii$
region originates primarily from polycyclic aromatic hydrocarbon (PAH)
molecules surrounding the extended 22$\mu$m emission from hot dust,
which is coincident with the ionized gas traced by radio continuum
emission \citep[e.g., see][]{dsa+10,hrds11} and RRLs.
The {\it WISE} catalog has more than 8\,400 entries of the Galactic $\hii$
regions and candidates, and lists their parameters of coordinates,
radius, RRL and/or molecular line velocity if available from
literature. This catalog is still being updated, and the
most recent version has been released in December 2020.

\begin{figure}[H]
\centering \includegraphics[width=0.35\textwidth]{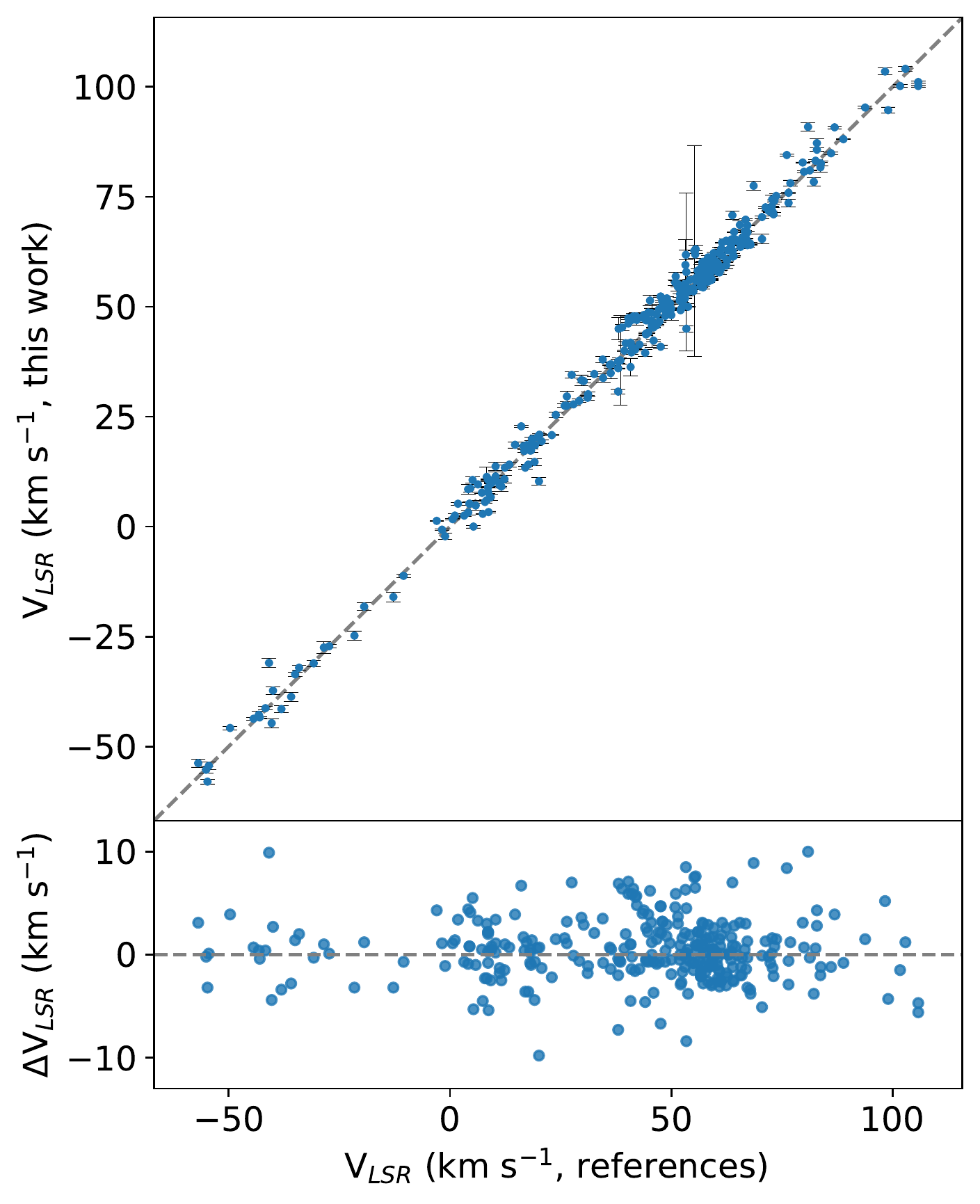}
\caption{A comparison of the FAST-measured V$_{\rm LSR}$ of 312 RRL
  components for the 302 known $\hii$ regions in the Galaxy area of
  $33\fdg05 \leqslant l \leqslant 54\fdg$95 and $|b| \leqslant
  1\fdg95 $ with the values in references. The dashed line is the
  equal line. 
  The differences, typically within a few km~s$^{-1}$ are shown in the lower sub-panel.}
\label{comvlsr}
\end{figure}

In the sky area of $33\fdg05 \leqslant l \leqslant 54\fdg$95 and
$|b| \leqslant 1\fdg95$, excluding the edges of images shown in
Figure~\ref{intall}, there are 1038 discrete sources in the {\it WISE}
catalog for $\hii$ regions and candidates. The velocities of
333 RRLs, V$_{\rm LSR}$, have been detected and reported for 322
$\hii$ regions, since some $\hii$ regions have two or three velocity
components.
Among them, 312 RRLs ($\sim$94\%) for 302 $\hii$ regions have been
detected from the piggyback spectral line data of the GPPS survey, as
marked by solid line circles in Figure~\ref{hii-distribution}.

Examples of line profiles for 4 $\hii$ regions are shown in
Figure~\ref{hiiexample}. In our RRL detection for some $\hii$ regions,
such as G34.026$-$0.058 in Figure~\ref{hiiexample}, some new RRL
components are often detected from the piggyback spectral data of the
GPPS survey but were not listed in the references.
They are probably produced from foreground or background diffuse
ionized gas along the line of sight towards an $\hii$ region, and may
be missed in the traditional ``on-off'' subtraction procedure. The
line processing method for the FAST GPPS survey piggyback data has the
advantage to detect weak and extended emission from the Galactic
ionized gas.

\begin{figure*}[t]
  \centering
  \includegraphics[width=0.8\textwidth]{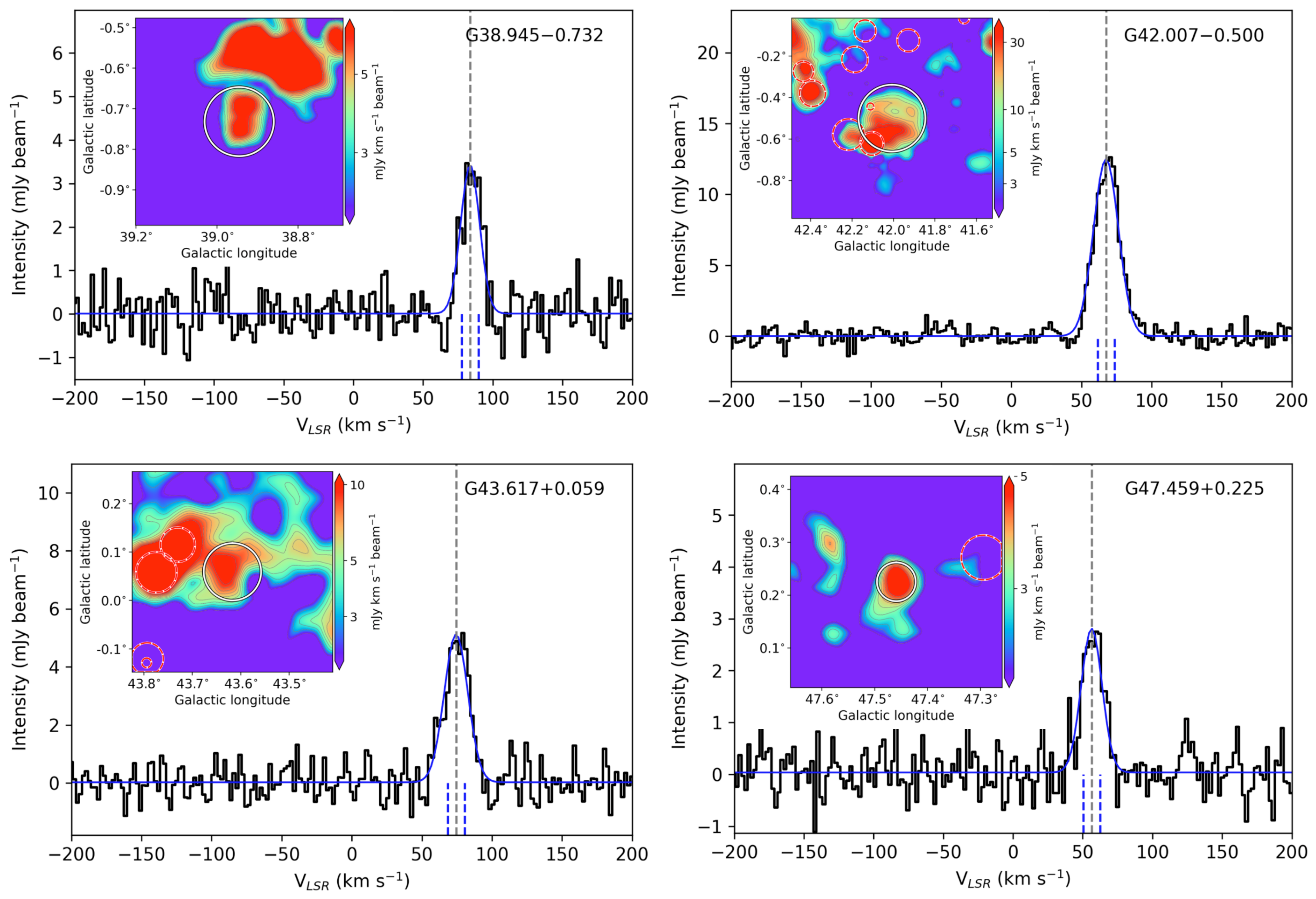}
   \caption{Same as Figure~\ref{hiiexample} but for examples of newly
     detected RRL spectra towards the candidates of $\hii$ regions
     listed in the latest version of the {\it WISE} catalog
     \citep[][]{wise14}. }
   \label{hiican}
\end{figure*}

Note, however, that 21 RRL components from known $\hii$ regions
reported in the references are not detected in this work: the
ultracompact $\hii$ region G33.134$-$0.093, G33.142$-$0.087 and
G41.740+0.096 observed by \citet{araya02} with the Arecibo at
4.874~GHz (H$110\alpha$); the $\hii$ region G34.197$-$0.593,
G34.198$-$0.591, G37.754+0.560, G37.820+0.414, G38.875+0.308, and
G39.883$-$0.346 measured by \citet{hrds11} with the GBT at $X$ band
(9~GHz); the distant $\hii$ region (V$_{\rm LSR}$~$<$~0.0~km~s$^{-1}$)
G39.183$-$1.422, G41.810+1.503, G42.058+1.002, G43.968+0.993,
G44.000+0.979, G52.002+1.602, G54.094+1.748, and G54.491+1.579
detected by \citet{hrds2015a} with the GBT at $X$ band (9~GHz);
G41.659$-$0.019 measured by \citet{arecibohii12} with the Arecibo
telescope at $X$ band (9~GHz); the ultracompact $\hii$ region
G43.793$-$0.121 and G43.794$-$0.129 observed by \citet{was03} with the
Arecibo at 4.874~GHz (H$110\alpha$); the V$_{\rm LSR} =
18.2$~km~s$^{-1}$ component of G45.883$-$0.087, detected by
\citet{hrds18} with the GBT at $C$ band (4$-$8~GHz). These RRL
components were mostly detected at a higher frequency and are probably
too weak or too compact at $L$ band to be detected in the GPPS
observations. It is also possible that the most compact $\hii$ regions
are optically thick at the observation frequencies of the FAST GPPS
survey.

The line profiles are fitted with Gaussian component(s) to derive the
V$_{\rm LSR}$ velocities and other parameters, as listed in
Table~\ref{knownhii}. We compare the FAST-obtained RRL velocities of
the 312 line components with the corresponding values in the
references, and they are very consistent as shown in
Figure~\ref{comvlsr}.

\begin{table*}[!t]
 \footnotesize
 \renewcommand\arraystretch{0.9}
\caption{H$n\alpha$ RRL parameters towards { 43} newly confirmed {\it WISE} $\hii$ regions }
\label{newhii}  
\tabcolsep 9pt 
\begin{tabular*}{\textwidth}{ccrrrrcrlr}

\toprule
\hline
   $\hii$ region & $l$          & $b$        &  Peak intensity    & V$_{\rm LSR}$    & FWHM         &  rms      & S/N   &  Mark  & $D_{\odot}$\\
                 & $(^\circ)$    & $(^\circ)$   &  (mJy~beam$^{-1}$)  & (km~s$^{-1}$)  & (km~s$^{-1}$)  & (mJy~beam$^{-1}$)  &      &     & (kpc) \\
   (1)           &  (2)         &   (3)      &    (4)             &   (5)         &  (6)          & (7)               &  (8) & (9) & (10)\\
   \hline
   G33.307$-$0.150 & 33.307 & $-$0.150 &      3.7 $\pm$      0.4 &     54.5 $\pm$      0.7 &     14.7 $\pm$      1.7 &     0.57 &      6.6 &  far  &  10.5$^{+0.4}_{-0.3}$ \\
   G33.487+0.159   & 33.487 & 0.159    &      1.9 $\pm$      0.2 &     27.7 $\pm$      1.1 &     18.3 $\pm$      2.6 &     0.53 &      3.6 &  far  &  11.9$^{+0.4}_{-0.4}$ \\
   G33.626+0.143   & 33.626 & 0.143    &      2.2 $\pm$      0.6 &     27.8 $\pm$      1.3 &      9.1 $\pm$      3.1 &     0.61 &      3.6 &  far  &  11.9$^{+0.4}_{-0.4}$ \\
   G33.636+0.174   & 33.636 & 0.174    &      3.0 $\pm$      0.2 &     31.0 $\pm$      0.7 &     16.9 $\pm$      1.6 &     0.32 &      9.6 &  far  &  11.7$^{+0.4}_{-0.4}$ \\
   G33.735$-$0.315 & 33.735 & $-$0.315 &      8.1 $\pm$      0.4 &     50.7 $\pm$      0.4 &     18.5 $\pm$      1.0 &     0.80 &     10.2 &  far  &  10.7$^{+0.4}_{-0.4}$ \\
   G33.818$-$0.086 & 33.818 & $-$0.086 &      1.4 $\pm$      0.5 &     31.2 $\pm$      0.9 &      4.9 $\pm$      2.1 &     0.31 &      4.6 &  far  &  11.7$^{+0.4}_{-0.4}$ \\
   G34.050$-$0.297 & 34.050 & $-$0.297 &      1.1 $\pm$      0.1 &     49.8 $\pm$      1.5 &     22.5 $\pm$      3.6 &     0.30 &      3.6 &  far  &  10.7$^{+0.4}_{-0.4}$ \\
   G34.131$-$0.173 & 34.131 & $-$0.173 &     11.7 $\pm$      0.4 &     48.3 $\pm$      0.6 &     15.7 $\pm$      1.2 &     0.71 &     16.5 &  far  &  10.7$^{+0.4}_{-0.4}$ \\
   G34.259+0.015   & 34.259 & 0.015    &      2.2 $\pm$      0.6 &  $-$33.8 $\pm$      0.9 &      6.9 $\pm$      2.1 &     0.48 &      4.5 &  out  &  16.5$^{+0.8}_{-0.7}$ \\
   G35.809+0.262   & 35.809 & 0.262    &      1.0 $\pm$      0.1 &     61.5 $\pm$      1.1 &     18.4 $\pm$      2.7 &     0.22 &      4.5 &  far  &   9.7$^{+0.4}_{-0.4}$ \\
   G36.106+0.146   & 36.106 & 0.146    &      9.9 $\pm$      1.2 &     92.8 $\pm$      0.7 &     11.6 $\pm$      1.7 &     2.62 &      3.8 &  ---      &      ---                \\
   G36.495+0.062   & 36.495 & 0.062    &      7.8 $\pm$      0.3 &     69.9 $\pm$      0.4 &     23.5 $\pm$      1.0 &     0.54 &     14.4 &  far  &   9.1$^{+0.4}_{-0.4}$ \\
   G36.738$-$0.301 & 36.738 & $-$0.301 &      7.7 $\pm$      0.4 &     80.7 $\pm$      0.3 &     11.5 $\pm$      0.7 &     0.58 &     13.3 &  near &   4.8$^{+0.6}_{-0.5}$ \\
   G37.237+0.741   & 37.237 & 0.741    &     11.8 $\pm$      0.3 &     42.2 $\pm$      0.2 &     16.4 $\pm$      0.4 &     0.63 &     18.7 &  far  &  10.5$^{+0.4}_{-0.4}$ \\
   G37.643+1.193   & 37.643 & 1.193    &      2.7 $\pm$      0.2 &     37.8 $\pm$      0.7 &     17.2 $\pm$      1.6 &     0.47 &      5.8 &  near &   2.3$^{+0.4}_{-0.4}$ \\
   G38.901$-$0.443 & 38.901 & $-$0.443 &      1.2 $\pm$      0.3 &     46.7 $\pm$      1.7 &     15.8 $\pm$      4.0 &     0.33 &      3.7 &  near &   2.9$^{+0.4}_{-0.4}$ \\
   G38.945$-$0.732 & 38.945 & $-$0.732 &      3.4 $\pm$      0.3 &     83.7 $\pm$      0.6 &     16.1 $\pm$      1.4 &     0.53 &      6.5 &  ---   &      ---          \\
   G39.214+0.202   & 39.214 & 0.202    &      0.8 $\pm$      0.1 &  $-$25.5 $\pm$      1.2 &     16.8 $\pm$      2.8 &     0.24 &      3.5 &  out  &  14.6$^{+0.6}_{-0.6}$ \\
   G39.462$-$0.273 & 39.462 & $-$0.273 &     17.9 $\pm$      0.2 &     63.8 $\pm$      0.2 &     25.0 $\pm$      0.4 &     0.50 &     35.7 &  far  &   8.8$^{+0.4}_{-0.5}$ \\
   G39.549$-$0.427 & 39.549 & $-$0.427 &      1.8 $\pm$      0.1 &     17.7 $\pm$      0.9 &     28.7 $\pm$      2.1 &     0.24 &      7.4 &  far  &  11.5$^{+0.4}_{-0.4}$ \\
   G40.100$-$1.409 & 40.100 & $-$1.409 &      9.5 $\pm$      0.5 &     48.3 $\pm$      0.4 &     16.4 $\pm$      1.0 &     1.07 &      8.8 &  near &   3.0$^{+0.4}_{-0.4}$ \\
   G40.306$-$0.432 & 40.306 & $-$0.432 &      1.8 $\pm$      0.1 &     68.8 $\pm$      1.0 &     24.4 $\pm$      2.3 &     0.32 &      5.5 &  near &   4.3$^{+0.6}_{-0.5}$ \\
   G40.399+0.683   & 40.399 & 0.683    &      2.5 $\pm$      0.3 &     10.0 $\pm$      0.6 &     12.3 $\pm$      1.5 &     0.50 &      5.0 &  near &   0.6$^{+0.5}_{-0.5}$ \\
   G40.620$-$0.650 & 40.620 & $-$0.650 &      2.4 $\pm$      0.2 &     70.2 $\pm$      0.8 &     20.0 $\pm$      1.9 &     0.36 &      6.8 &  ---  &       ---               \\
   G41.948$-$0.261 & 41.948 & $-$0.261 &      2.5 $\pm$      0.1 &     14.0 $\pm$      0.6 &     19.3 $\pm$      1.3 &     0.30 &      8.1 &  far  &  11.3$^{+0.4}_{-0.4}$ \\
   G41.950+0.543   & 41.950 & 0.543    &      6.3 $\pm$      0.2 &     24.8 $\pm$      0.3 &     17.3 $\pm$      0.7 &     0.42 &     15.1 &  far  &  10.6$^{+0.4}_{-0.4}$ \\
   G42.007$-$0.500 & 42.007 & $-$0.500 &     11.6 $\pm$      0.4 &     67.4 $\pm$      0.4 &     19.3 $\pm$      0.9 &     1.09 &     10.6 &  ---  &        ---              \\
   G42.228$-$0.066 & 42.228 & $-$0.066 &     15.2 $\pm$      0.2 &     53.9 $\pm$      0.2 &     20.3 $\pm$      0.4 &     0.53 &     28.5 &  near &   3.4$^{+0.5}_{-0.5}$ \\
   G43.617+0.059   & 43.617 & 0.059    &      5.1 $\pm$      0.3 &     74.4 $\pm$      0.5 &     19.9 $\pm$      1.2 &     0.44 &     11.5 &  tan  &   6.0$^{+1.5}_{-1.5}$ \\
   G43.936$-$0.831 & 43.936 & $-$0.831 &     13.8 $\pm$      0.2 &     55.3 $\pm$      0.2 &     21.2 $\pm$      0.4 &     0.45 &     30.6 &  near &   3.7$^{+0.6}_{-0.5}$ \\
   G44.586+0.363   & 44.586 & 0.363    &      2.0 $\pm$      0.1 &     17.3 $\pm$      0.5 &     14.5 $\pm$      1.2 &     0.27 &      7.3 &  far  &  10.5$^{+0.4}_{-0.4}$ \\
   G44.645+0.369   & 44.645 & 0.369    &      1.0 $\pm$      0.1 &     19.1 $\pm$      0.9 &     15.6 $\pm$      2.1 &     0.21 &      4.9 &  far  &  10.4$^{+0.4}_{-0.4}$ \\
   G45.191$-$0.485 & 45.191 & $-$0.485 &      6.5 $\pm$      0.2 &     67.2 $\pm$      0.4 &     26.3 $\pm$      1.0 &     0.55 &     11.8 &  ---  &        ---              \\
   G46.253$-$0.585 & 46.253 & $-$0.585 &      4.6 $\pm$      0.3 &     56.9 $\pm$      0.4 &     13.9 $\pm$      1.1 &     0.50 &      9.2 &  ---  &        ---              \\
   G46.792$-$0.264 & 46.792 & $-$0.264 &      4.7 $\pm$      0.3 &     50.5 $\pm$      0.6 &     19.2 $\pm$      1.5 &     0.73 &      6.4 &  far  &   7.7$^{+0.6}_{-0.6}$ \\
   G46.792+0.285   & 46.792 & 0.285    &      7.5 $\pm$      0.5 &  $-$52.8 $\pm$      0.7 &     23.0 $\pm$      1.7 &     1.33 &      5.6 &  out  &  15.5$^{+0.8}_{-0.7}$ \\
   G47.459+0.225   & 47.459 & 0.225    &      2.8 $\pm$      0.2 &     56.4 $\pm$      0.6 &     17.6 $\pm$      1.5 &     0.42 &      6.6 &  ---   &            ---          \\
   G49.695$-$0.042 & 49.695 & $-$0.042 &      3.0 $\pm$      0.1 &     52.1 $\pm$      0.5 &     25.0 $\pm$      1.3 &     0.34 &      8.7 &  ---  &            ---          \\
   G49.728+0.129   & 49.728 & 0.129    &      4.4 $\pm$      0.2 &     47.0 $\pm$      0.3 &     13.7 $\pm$      0.8 &     0.38 &     11.7 &  near &   3.6$^{+0.7}_{-0.7}$ \\
   G49.801$-$0.436 & 49.801 & $-$0.436 &      6.8 $\pm$      0.2 &     61.6 $\pm$      0.3 &     24.2 $\pm$      0.8 &     0.46 &     14.6 &  tan  &   5.4$^{+1.7}_{-1.7}$ \\
   G51.457$-$0.285 & 51.457 & $-$0.285 &      6.9 $\pm$      0.6 &     57.7 $\pm$      0.6 &     14.9 $\pm$      1.5 &     1.38 &      5.0 &  tan  &   5.2$^{+1.7}_{-1.7}$ \\
   G52.437+0.197   & 52.437 & 0.197    &      3.8 $\pm$      0.6 &     55.3 $\pm$      0.7 &      8.8 $\pm$      1.5 &     0.82 &      4.6 &  tan  &   5.1$^{+1.7}_{-1.7}$ \\
   G53.047$-$0.818 & 53.047 & $-$0.818 &      5.3 $\pm$      0.3 &     61.6 $\pm$      0.3 &     14.4 $\pm$      0.8 &     0.45 &     12.0 &  tan  &   5.0$^{+1.7}_{-1.7}$ \\
\hline
\bottomrule
\end{tabular*}
\begin{tablenotes}
\item[1] { Notes.} Column 1 is the source name; Cols. 2 and 3 are the
  Galactic longitude and the Galactic latitude given by the {\it WISE}
  catalogue of $\hii$ regions and candidates \citep[][]{wise14};
  Cols. 4, 5, and 6 are the fitted peak intensity, V$_{\rm LSR}$
  velocity and line width (FWHM) by this work, together with their
  1$\sigma$ uncertainties; Col. 7 is the rms noise of the RRL spectra
  for every object; Col. 8 is the signal-to-noise ratio of the RRL
  spectrum; Col. 9 is a mark for the kinematic distance ambiguity
  resolved by this work: near – the nearer kinematic distance is
  adopted; far – the farther kinematic distance is adopted; tan – the
  tracer is located at the tangential region and the distance to the
  tangent is adopted; out – the tracer is in the outer Galaxy outside
  the solar circle; Col. 10 lists the kinematic distance and its
  uncertainty calculated by using the program given by
  \citet[][]{rmb+14}.
\end{tablenotes}
\end{table*}

\subsection{New $\hii$ regions and distant star-forming regions}
\label{new}
In the concerned sky area of this paper, more than 710 candidates of
$\hii$ regions are listed in the {\it WISE} catalog
\citep[][]{wise14}. We extract the H$n\alpha$ RRL spectra towards these
candidates in order to verify their nature.

Considering the FAST beam of 3$'$, in order to obtain a reliable
detection of RRL toward an $\hii$ region candidate, we select
relatively isolated targets, not contaminated by nearby $\hii$ regions
or candidates inside a FAST beam. In addition, to avoid the confusion
of the RRLs from diffuse gas, the RRLs are obtained by spectra of the
target subtracted by the mean reference spectrum towards some selected
positions around the target.
Then, the line profile is fitted. If the velocity-integrated intensity
map within the $\pm$6~km~s$^{-1}$ around the fitted line center has an
isolated feature, coincident with the position of the $\hii$ region
candidate inside the diameter, we take it as a good detection.

We get H$n\alpha$ RRLs of 43 $\hii$ region candidates
detected in the sensitive spectral data of the FAST GPPS survey,
as marked in Figure~\ref{hii-distribution}.  The line parameters for the 43 $\hii$ regions are
listed in Table~\ref{newhii}, and all of them have a signal-to-noise
ratio greater than 3.5$\sigma$ at the spectral resolution of
2.2~km~s$^{-1}$. Some examples of the line
profiles with a velocity-integrated intensity map are shown in
Figure~\ref{hiican}. 

\begin{figure}[H]
  \centering
   \includegraphics[width=0.43\textwidth]{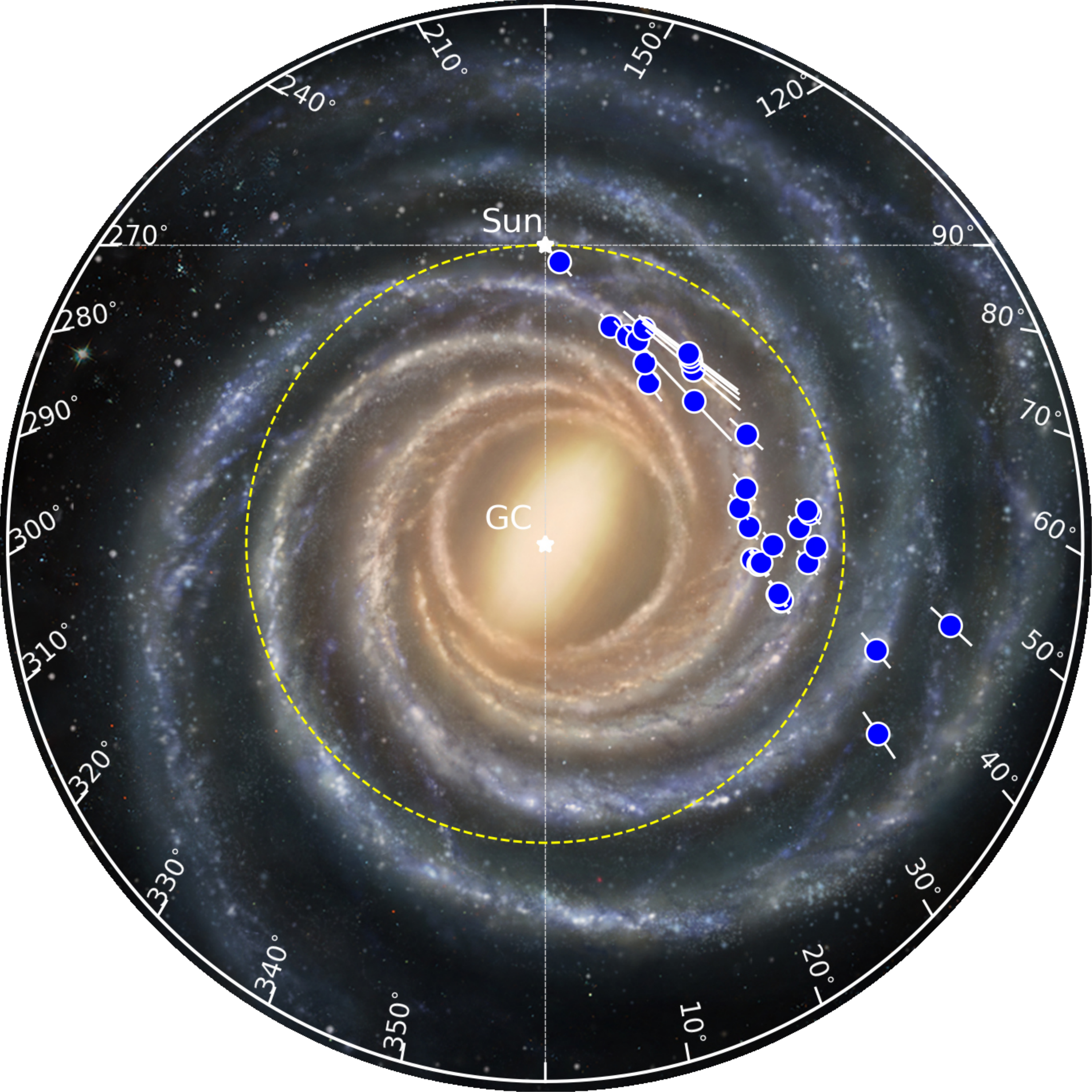}\\
   \caption{Distribution of 35 newly confirmed $\hii$ regions (see
     Table~\ref{newhii}) projected onto the Galactic plane with
     position error bars. Their kinematic distances are calculated by
     using the method described in \citet{rmb+14}. The background is a
     concept map of the spiral structure of the Milky Way (credited by
     Xing-Wu Zheng \& Mark Reid, BeSSeL/NJU/CFA). Some $\hii$ regions
     in the very outer Galaxy have been detected.}
   \label{distant}
\end{figure}

The kinematic distances for 35 of the 43 $\hii$ regions are estimated
after resolving the kinematic distance ambiguity through a
combination of the $\hi$ emission/absorption method and the $\hi$
self-absorption method \citep[][]{hg19}. We adopt the program of
\citet{rmb+14} to calculate the kinematic distances by using the
Galactic dynamic model, taking the distance of the Sun to the Galactic
center $R_\odot$ being 8.34~kpc, the circular orbital speed at the Sun
$\Theta_0$ being 240~km~s$^{-1}$, and the solar motion
being $U_\odot = 10.7$ ~km~s$^{-1}$ (radially inward towards the
Galactic center), $V_\odot = 15.6$~km~s$^{-1}$ (in the direction of
Galactic rotation), and $W_\odot = 8.9$~km~s$^{-1}$ (vertically upward
towards the north Galactic pole) in J2000 \citep[][]{rmb+14}. The
distribution of these newly confirmed $\hii$ regions is presented in
Figure~\ref{distant}. They are primarily located in the
Sagittarius-Carina Arm, the Perseus Arm, and the Norma-Outer Arm. A
few of them seem to be related with the arm spur located between the
Sagittarius Arm and the Scutum Arm.

In the current survey area, at least seven extended RRL sources, e.g.,
G39.516+0.512 and G44.501+0.332, have been detected in the outer
Galaxy, i.e. their Galactocentric distance larger than the solar
circle in the first Galactic quadrant. At the kinematic distances
greater than about 12~kpc from the Sun, their physical sizes are
estimated to be 30$-$100~pc, and are among the largest ones comparable
to the star-forming complex W\,51 in the inner Galaxy. Therefore,
the piggyback spectral data of the FAST GPPS survey show its ability
to detect such large $\hii$ regions or star-forming complexes
previously not known in the very distant outer Galaxy, the locations
of which are valuable laboratories for studying the star formation
process in a lower gas density and lower metallicity environment than
in the inner Galaxy \citep[e.g.,][]{bw07,si21}.

\begin{figure*}[t]
  \centering \includegraphics[width=0.85\textwidth]{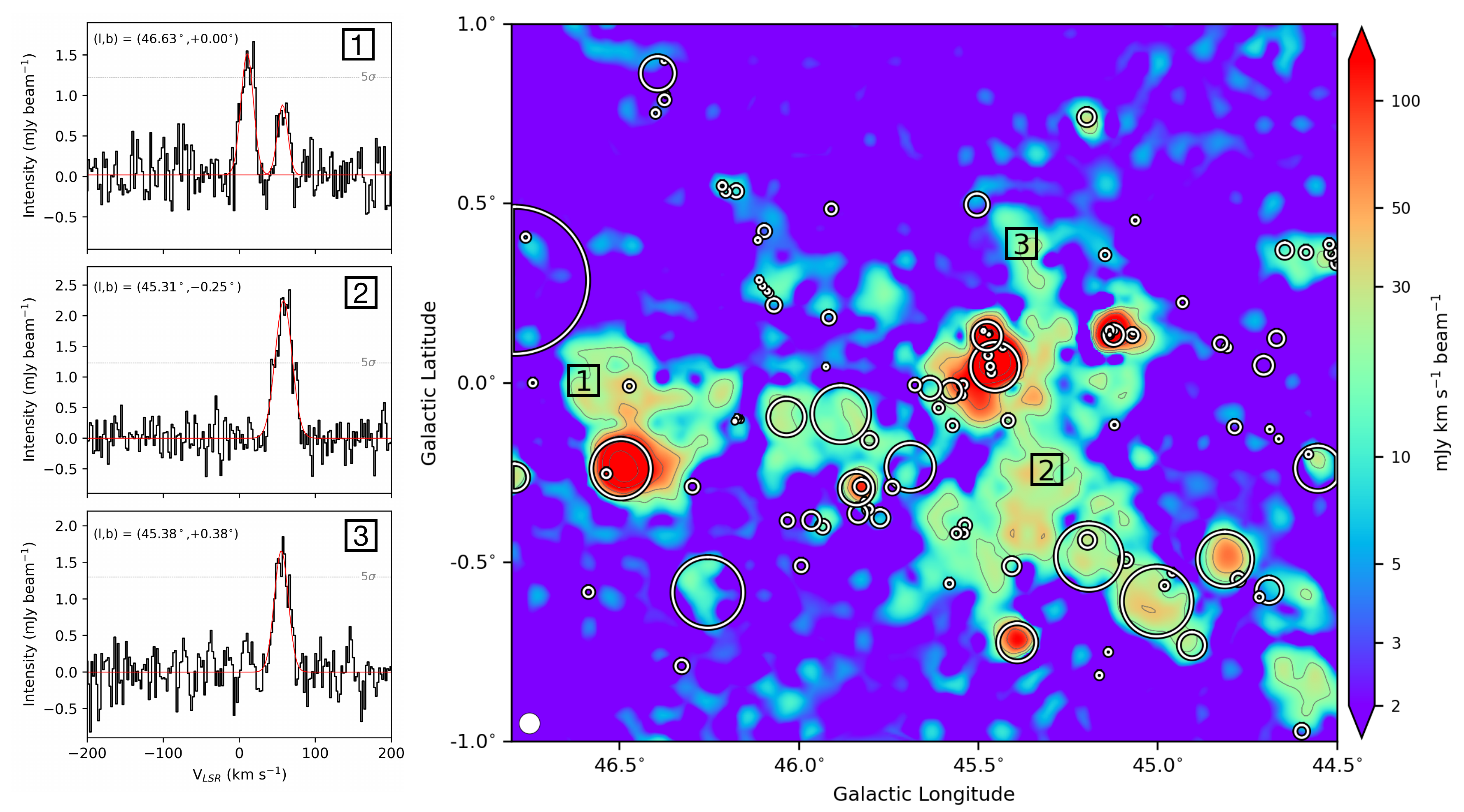}\\
   \caption{Examples of the H$n\alpha$ RRL spectra for the diffuse
     ionized gas ({\it left panels}) at positions marked in the
     velocity-integrated intensity map ({\it right panel}). The
     circles indicate the boundaries of the {\it WISE} $\hii$ regions
     and candidates, and the sizes correspond to source radii in the
     {\it WISE} catalog. Outside these discrete sources, the RRL
     emission comes primarily from diffuse ionized gas. The contour
     levels are the same as that of Figure~\ref{intall}. The white
     filled circle in the lower-left corner of the map indicates the
     beam size ($\sim3^\prime$) of FAST at $L$ band.}
   \label{diffuse}
\end{figure*}

\subsection{Diffuse ionized gas}

The diffuse ionized gas is widely distributed in the Milky Way. It
accounts for about 90\% of the total mass of ionized gas in the Milky
Way \citep[][]{rey91}, playing an important role in the recycling of
material in the Galactic ISM. The physical properties and origin of
diffuse ionized gas have not been well understood~\citep[][]{gdigs}.

The existence of diffuse ionized gas in the Milky Way was first
proposed by \citet{he63}. Though some diffuse ionized gas exists in
the form of the low-density envelopes around individual $\hii$ regions
\citep[e.g.,][]{gdigs,ana85}, most diffuse gas 
is far from well-defined $\hii$ regions. 

With the high sensitivity inherited in the snapshot observation mode,
the piggyback spectral line data of the FAST GPPS survey provide the
great advantage to detect the diffuse ionized gas in the Galactic
plane. As discussed in Sect.~\ref{results_sec}, the H$n\alpha$ data
products of the GPPS project are sensitive to emission measures down
to 200~cm$^{-6}$~pc if a 3$\sigma$ detection limit is required.
As shown in Figure~\ref{diffuse} for the Galactic longitude range of
$44\fdg5 \leqslant l \leqslant 46\fdg$8 and the Galactic latitude
range of $|b| \leqslant 1\fdg0$, we can detect the RRLs beyond the
known $\hii$ regions and $\hii$ region candidates
\citep[][]{wise14}. By using the {\it WISE} 12~$\mu$m and 22~$\mu$m
survey images, the $\hii$ region
boundaries~\citep[e.g.,][]{wise14,dsa+10} can be outlined.
Most of the RRL emissions outside the known $\hii$ regions and
candidates, if not all, are from diffuse ionized gas.
The emission measures for the four RRL components shown in the left
panels of Figure~\ref{diffuse} are about 1100, 700, 1800 and
1300~cm$^{-6}$~pc, respectively.
The features of diffuse ionized gas are prominent and even more
complex in some other regions, such as around the star-forming complex
W\,51, W\,49, and W\,43.

\begin{figure*}[!t] 
  \centering
\includegraphics[width=0.33\textwidth]{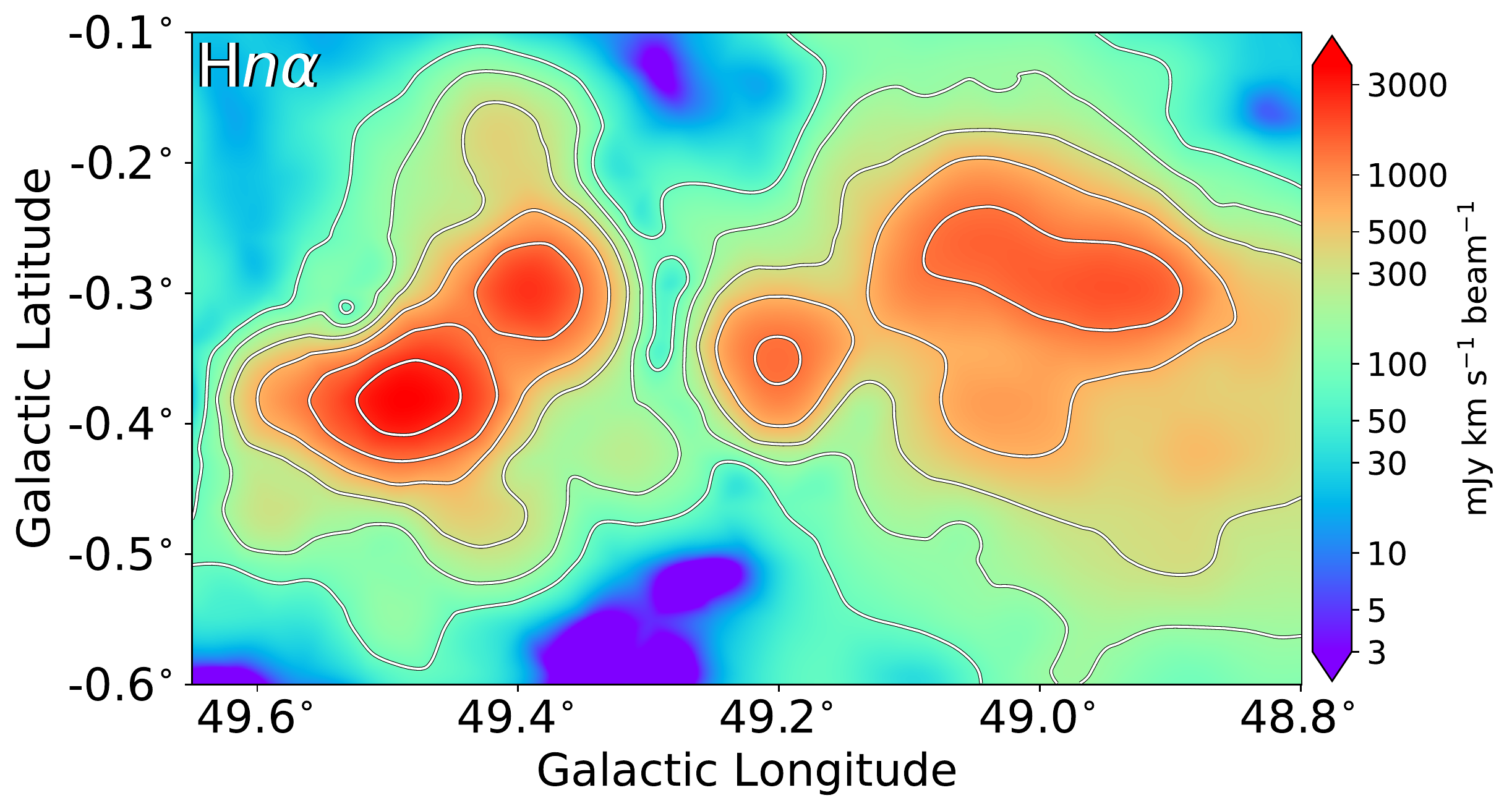}
\includegraphics[width=0.33\textwidth]{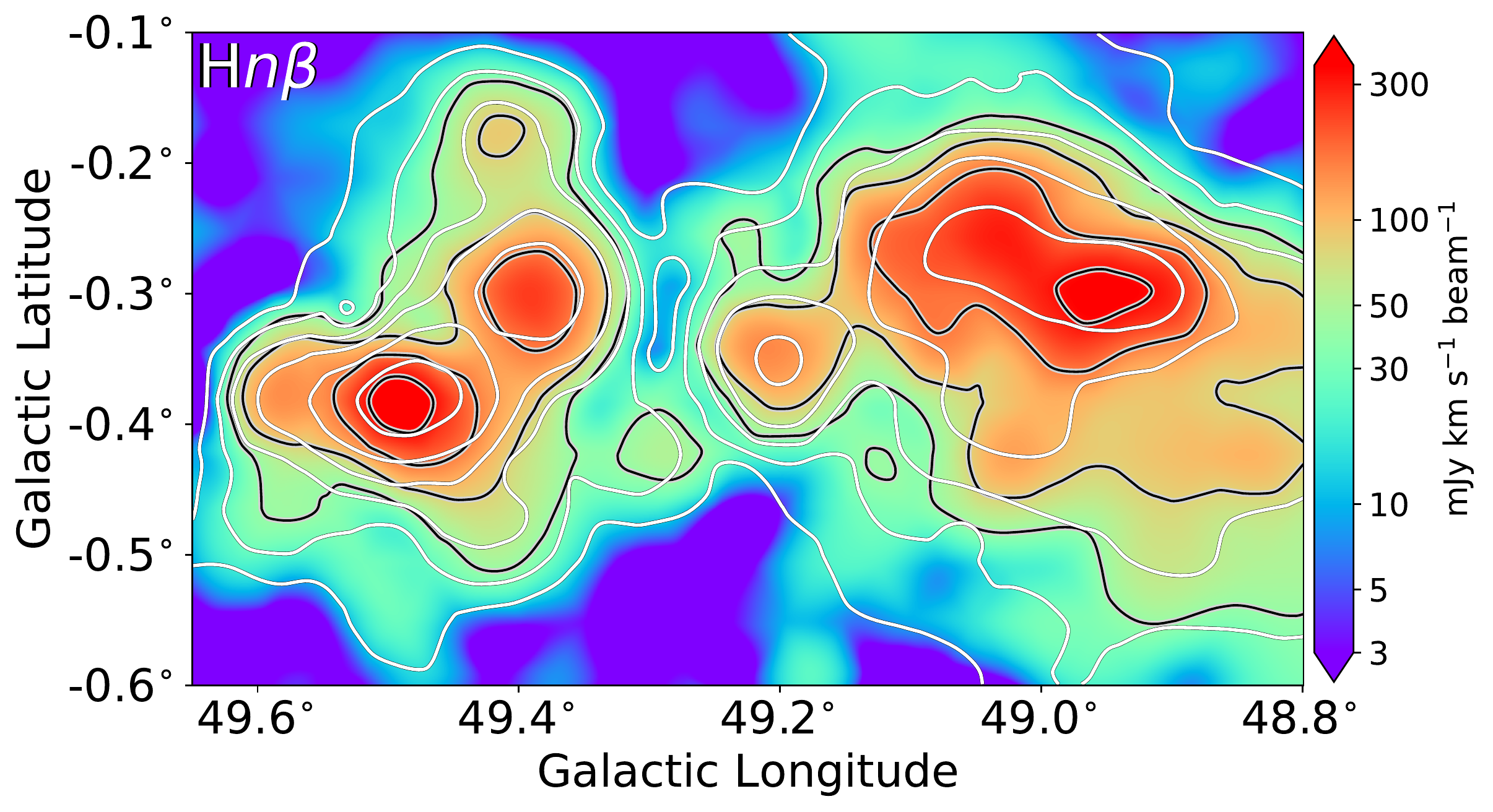}
\includegraphics[width=0.33\textwidth]{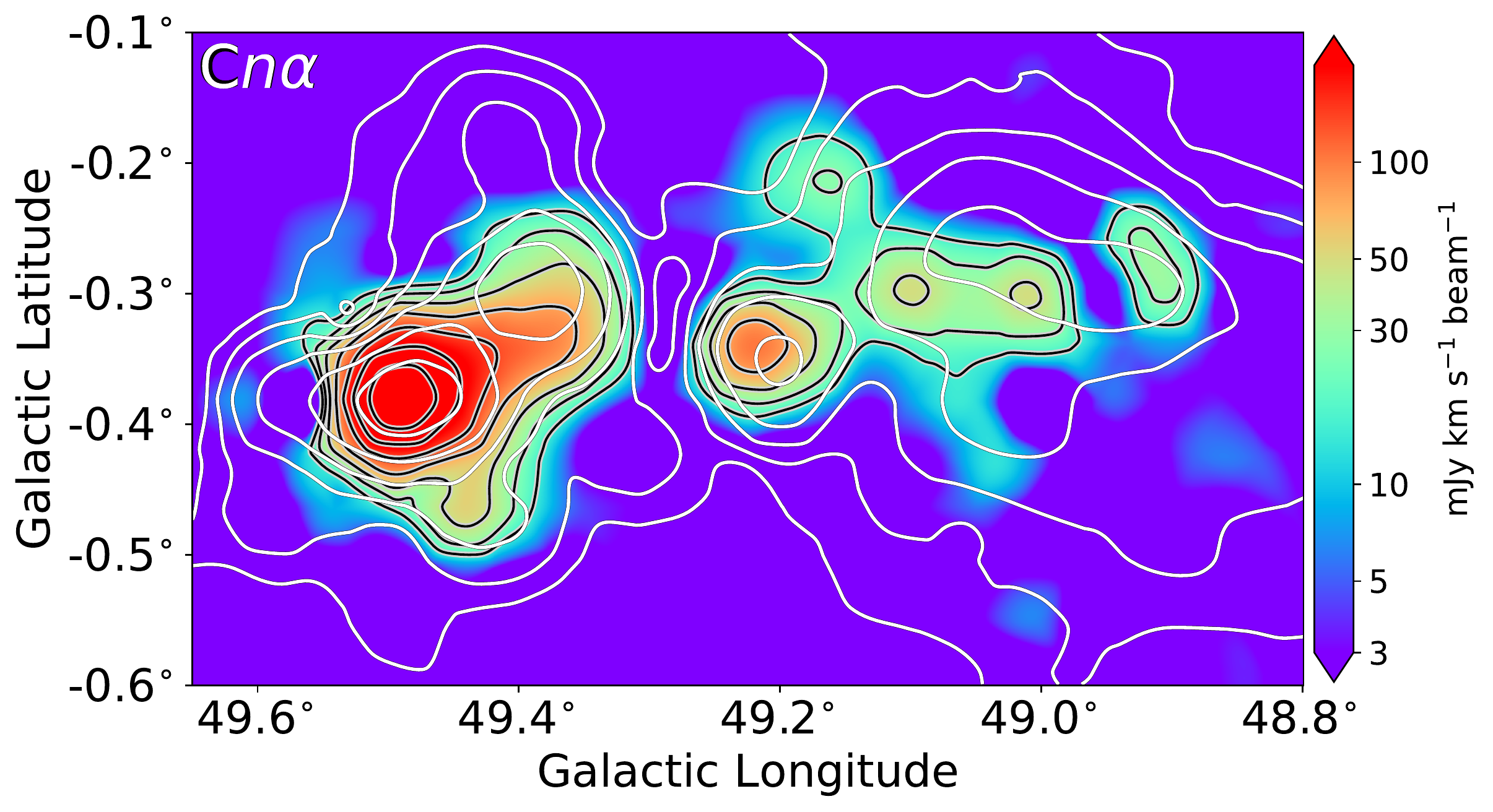}
   \caption{Velocity-integrated intensity map of the H$n\alpha$ RRLs
     ({\it left panel}), the H$n\beta$ RRLs ({\it middle panel}) and
     C$n\alpha$ RRLs ({\it right panel}) for the star-forming complex
     W\,51 obtained by the piggyback spectral data of the FAST GPPS
     survey. The contours in the {\it left} panel at levels of
     $2^n\times$~5~mJy~km~s$^{-1}$~beam$^{-1}$ with $n=$~4 to 10 stand
     for the velocity-integrated intensity of the H$n\alpha$ RRLs,
     which are also plotted as white contours in the {\it middle} and
     {\it right} panels for comparison.  The contour levels are at
     levels of $2^n\times$~5~mJy~km~s$^{-1}$~beam$^{-1}$ with $n=$~3
     to 6 for the H$n\beta$ RRLs, and at levels of
     $\sqrt{3}^n\times$~5~mJy~km~s$^{-1}$~beam$^{-1}$ with $n=$~2 to
     11 for the C$n\alpha$ RRLs.  }
   \label{carbon}
\end{figure*}

\section{Conclusions and discussions}
\label{sect:conclusion}

The full FAST GPPS survey~\citep{gpps} aims to hunter for faint
pulsars in the Galactic plane with $\sim 30^\circ \lesssim l \lesssim
\sim98^\circ$, $\sim 148^\circ \lesssim l \lesssim \sim 216^\circ$,
and $|b| < 10^\circ$ that is visible to FAST. The piggyback spectral
data recorded simultaneously during the survey observations are
valuable resources for the RRL detection in the frequency range of
1000 $-$ 1500~MHz for the Galactic $\hii$ regions and diffuse ionized
gas in the interstellar medium, which will improve our knowledge on
the distribution, the origin and characteristics of the Galactic
ionized gas, the star formation process and spiral structure.

We process the data and present the results for the H$n\alpha$ RRLs
for a sky area of 88 square degrees in the inner Galaxy of $33^\circ
\leqslant l \leqslant 55^\circ$ and $|b| \leqslant 2\fdg0$. The
averaged H$n\alpha$ RRLs from the FAST snapshot mode observations with
an integration time of 5-minute for each beam reach a great
sensitivity of the typical rms noise of about 0.25~mJy~beam$^{-1}$ (or
6.3~mK~$T_B$) at a spectral resolution of 2.2~km~s$^{-1}$, currently
it is the most sensitive survey of RRLs at the $L$ band.

The RRL map of this sky region shows complex and abundant structural
features for discrete $\hii$ regions and diffuse ionized gas. Some
large star-forming complexes in the inner Galaxy, such as W\,51, W\,49
and W\,47 are prominent strong features in the map. The detected RRL
emission are primarily related to two major spiral arms and two arm
spurs. Among 322 known $\hii$ regions, about 94\% of them have been
detected, and the measured line velocities of averaged RRLs are in
good agreement with the values in literature.
The undetected $\hii$ regions are either too weak or too compact at
$L$ band to be detected in the GPPS observations.
We detected RRLs toward 43 $\hii$ region candidates, confirming their
nature as being $\hii$ regions, increasing the sample by a factor of
about 1/8. The spectral line data of the FAST GPPS survey have great
advantage to detect and resolve large distant $\hii$ regions or
star-forming complexes even in the outer Galaxy, and can reveal
diffuse ionized gas widely spread in the Galactic plane, which has not
been well mapped previously.

As shown in Table~\ref{rrls}, beside the H$n\alpha$ RRLs presented
above, other kinds of RRLs have also been recorded in the observation
band of the FAST GPPS survey, including 30 H$n\beta$, 34
H$n\gamma$, 23 C$n\alpha$ and 23 He$n\alpha$ RRLs.
The H$n\beta$, H$n\gamma$ and He$n\alpha$ RRLs are generally from the
same sources as that of the H$n\alpha$ RRLs, but have much weaker line
strengths.  
The FAST GPPS spectral line data can also provide a sensitive
detection for the RRLs of H$n\beta$ and H$n\gamma$ at $L$ band, which
have seldom been studied previously.

We processed the data for H$n\beta$ and C$n\alpha$ RRLs for a small
sample area around the star-forming complex W\,51, { and} show the
results in Figure~\ref{carbon}.
The H$n\beta$/H$n\alpha$ and/or H$n\gamma$/H$n\alpha$ ratios will
provide useful information about the physics of ionized gas, e.g., the
situations of local thermodynamic equilibrium (LTE) or non-LTE.
The Carbon RRLs are either from the photo-dissociation regions at the
interface between $\hii$ regions and their surrounded molecular
clouds, the boundary between molecular clouds and the diffuse ISM, or
from the regions within atomic $\hi$ clouds \citep[e.g.,][]{gs02}.
Compared with the map of the H$n\alpha$ RRLs, the velocity-integrated
intensity map of C$n\alpha$ RRLs in Figure~\ref{carbon} has { peaks
  clearly offset} from those of H$n\alpha$ RRLs, suggesting that the
C$n\alpha$ and H$n\alpha$ RRLs originate from different regions inside
the star formation complex. The data products of C$n\alpha$ RRLs as
well as other kinds of RRLs are therefore also useful byproducts of
the GPPS survey, which are under processing now and will be published
in the future.
The ionization energy for helium is 24.6~eV, much higher than
that of hydrogen (13.6~eV). 
survey~\citep{gpps} has the capability to detect the weak He$n\alpha$
RRLs from a number of Galactic sources, which is important for
understanding the ionization sources by analyzing the He/H ratio and
the He/H elemental abundance ratio \citep[e.g.,][]{rrl80,gs02}.

As the most sensitive survey for RRLs at $L$ band, together with good
spatial resolution, spectral resolution and large sky coverage, the
piggyback spectral data of the FAST GPPS survey will largely promote
our understanding of the properties of Galactic ionized gas.

\vspace{-2mm}
\section*{Data availability}

Original FAST GPPS survey data, including the piggyback recorded
spectral line data, are released one year after observations,
according to the FAST data release policy. All processed RRL data as
presented in this paper are available on the project web-page:
{\color{blue}http://zmtt.bao.ac.cn/MilkyWayFAST/}.

\vspace{+8mm}
%
{\footnotesize \it
This work is supported by the National Natural Science Foundation of
China (NSFC) No. 11988101, 11933011, 11833009, 12133004, 12003044, the
National Key R\&D Program of China (Grant No. 2017YFA0402701), the Key
Research Program of the Chinese Academy of Sciences (Grant
No. QYZDJ-SSW-SLH021) and the National SKA program of China
No. 2022SKA0120103. LGH and CW thank the support from the Youth
Innovation Promotion Association CAS. XYG acknowledges the financial
support by the CAS-NWO cooperation programme (Grant No. GJHZ1865) and
by the NSFC No. U1831103. The work is based on the piggyback spectral
data simultaneously recorded during the observations of the GPPS
survey, which is one of the five key projects carried out by using the
Five-hundred-meter Aperture Spherical radio Telescope (FAST). FAST is
a Chinese national mega-science facility built and operated by the
National Astronomical Observatories, Chinese Academy of Sciences.
}

{\small \setlength{\baselineskip}{-1pt}
\bibliographystyle{raa}
\bibliography{bibtex}
}
\end{multicols}

\end{document}